\pdfoutput=1

\documentclass[11pt]{article}
\RequirePackage[OT1]{fontenc}
\RequirePackage{amsthm,amsmath}
\RequirePackage[numbers]{natbib}
\RequirePackage[colorlinks,citecolor=blue,urlcolor=blue]{hyperref}

\usepackage[utf8]{inputenc} 
\usepackage[T1]{fontenc}    
\usepackage[colorlinks]{hyperref} 
\usepackage{url}            
\usepackage{booktabs}       
\usepackage{amsfonts}       
\usepackage{amsmath}
\usepackage{nicefrac}       
\usepackage{microtype}      
\usepackage{color}
\usepackage{bbm}
\usepackage{amssymb, enumerate}
\usepackage{makecell}
\usepackage{wrapfig}
\usepackage{extarrows}
\usepackage{arydshln}

\usepackage{caption}
\usepackage{multirow}

\usepackage{amsmath,amsfonts,amsthm, amssymb, enumerate}
\usepackage{lipsum}
\usepackage{caption}
\usepackage{amssymb,graphicx,subfigure}
\usepackage[bottom]{footmisc}
\usepackage[dvipsnames]{xcolor}
\usepackage{array,multirow}
\usepackage{enumitem}

\usepackage{microtype}
\usepackage{graphicx}
\usepackage{subfigure}
\usepackage{booktabs} 
 
\usepackage{graphicx}
\usepackage{amsmath}
\usepackage{amssymb}
\usepackage{booktabs}  
\usepackage{times}
\usepackage{epsfig}
\usepackage{graphicx}
\usepackage{amsfonts, amsthm, enumerate} 
\usepackage[bottom]{footmisc} 
\usepackage{enumitem}
\usepackage{setspace}
\usepackage{wrapfig}
\usepackage{extarrows}
\usepackage{makecell}
\usepackage{bbm}  
\usepackage{textcomp}
\usepackage{xcolor} 
\usepackage{caption}

\newcommand{\argmin}{\operatornamewithlimits{argmin}}
\newcommand{\argmax}{\operatornamewithlimits{argmax}}

\usepackage[font=small,labelfont=bf]{caption}

\usepackage{algorithm,algpseudocode}
\algnewcommand\TR{\item[{\textbf{Training phase}}]}
\algnewcommand\TE{\item[{\textbf{Test phase}}]}
\algnewcommand\Input{\item[{{Input:}}]}
\algnewcommand\Output{\item[{{Output:}}]}
\algnewcommand\Initialize{\item[{{Initialize:}}]}
\algnewcommand{\return}[1]{
	\State \textbf{return:}
	\Statex \hspace*{\algorithmicindent}\parbox[t]{.8\linewidth}{\raggedright #1}
}


\usepackage{setspace}
\begin{document}
	\title{3D unsupervised anomaly detection and localization through virtual multi-view projection and reconstruction: Clinical validation on low-dose chest computed tomography}
	\date{}
	\author{
		Kyung-Su Kim$^{1,2}$\thanks{Equal contribution}\,\,\thanks{Corresponding author: Kyung-Su Kim (kskim.doc@gmail.com) and Myung Jin Chung (mj1.chung@samsung.com)}, Seong Je Oh$^{3}$\footnotemark[1], Ju Hwan Lee$^{3}$, Myung Jin Chung$^{1,4}$\footnotemark[2]\\ 
		{\small $^{1}$Medical AI Research Center, Research Institute for Future Medicine, Samsung Medical Center, Seoul, Korea}\\
		{\small $^{2}$Department of Data Convergence and Future Medicine, Sungkyunkwan University School of Medicine, Seoul, Korea}\\
		{\small $^{3}$Department of Health Sciences and Technology, SAIHST, Sungkyunkwan University, Seoul, Korea}\\
		{\small $^{4}$Department of Radiology, Samsung Medical Center, Sungkyunkwan University School of Medicine, Seoul, Korea}
	}

	\maketitle 
	
\begin{abstract}
Computer-aided diagnosis for low-dose computed tomography (CT) based on deep learning has recently attracted attention as a first-line automatic testing tool because of its high accuracy and low radiation exposure. However, existing methods rely on supervised learning, imposing an additional burden to doctors for collecting disease data or annotating spatial labels for network training, consequently hindering their implementation. We propose a method based on a deep neural network for computer-aided diagnosis called virtual multi-view projection and reconstruction for unsupervised anomaly detection. Presumably, this is the first method that only requires data from healthy patients for training to identify three-dimensional (3D) regions containing any anomalies. The method has three key components. Unlike existing computer-aided diagnosis tools that use conventional CT slices as the network input, our method 1) improves the recognition of 3D lung structures by virtually projecting an extracted 3D lung region to obtain two-dimensional (2D) images from diverse views to serve as network inputs, 2) accommodates the input diversity gain for accurate anomaly detection, and 3) achieves 3D anomaly/disease localization through a novel 3D map restoration method using multiple 2D anomaly maps. The proposed method based on unsupervised learning improves the patient-level anomaly detection by 10$\%$ (area under the curve, 0.959) compared with a gold standard based on supervised learning (area under the curve, 0.848), and it localizes the anomaly region with 93\% accuracy, demonstrating its high performance.
\end{abstract}
	
\section{Introduction}

\begin{figure*}[t]
\centerline{\includegraphics[width=0.8\textwidth]{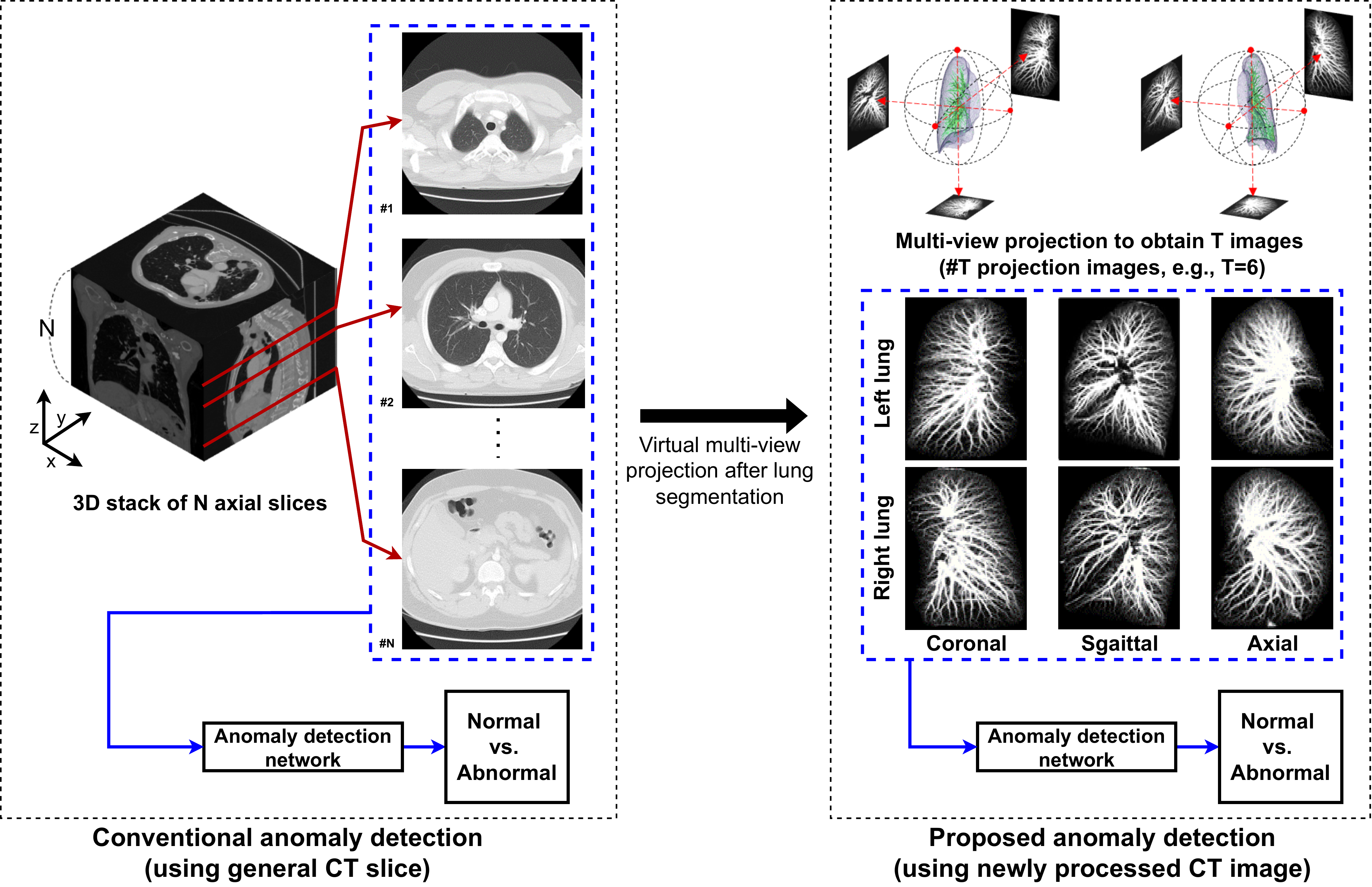}} 
\caption{Proposed anomaly detection compared with baseline. Our method uses virtual multi-view projected images as input for diagnosis, whereas the baseline uses conventional CT slices.}
\label{fig:proposed_img}
\end{figure*}

\subsection{Importance of low-dose computed tomography}

\,\,\,\,\,\,\,\, In radiology, computed tomography (CT) is widely used for precise diagnosis of internal diseases in the chest and abdomen. In particular, CT is more sensitive than simple chest radiography for diagnosis of lung diseases \citep{stephanie2020determinants}. In addition, compared with magnetic resonance imaging, CT provides higher image resolution at a lower examination price \citep{hong2000high,kim2002usefulness}. However, as normal-dose CT generally exposes patients to high levels of radiation \citep{brenner2007computed,mancini2010impact,brenner2012cancer}, 
few CT scans can be acquired due to ethical concerns. Alternatively, low-dose CT (LDCT) is attracting attention for examination {\citep{larke2011estimated}}, as it reduces the radiation exposure in approximately three times compared with the normal-dose examination, improving safety from patient exposure while allowing to detect microscopic diseases that cannot be seen in normal X-ray images. Thus, LDCT is widely used as the first-line test for detecting lung anomalies with high sensitivity. For instance, it allows detecting small lung diseases, such as small lung cancer that cannot be observed in X-ray images, and provides their three-dimensional (3D) location, being widely used for initial screening of lung cancer in high-risk groups (e.g., smokers and elders). However, LDCT may present a higher false-positive rate than normal-dose CT in diagnosing lung diseases, as it may lead to increased noise, artifacts, or non-detection owing to unsatisfactory image reconstruction given the low-dose imaging.

\subsection{Deep-learning-based diagnosis in LDCT} 
 
\,\,\,\,\,\,\,\, To improve diagnosis using LDCT, various studies have been aimed to reduce the false-positive rate by using automatic deep-learning-based computer-aided diagnosis in LDCT. Supervised learning (SL) methods for diagnosis using medical images allow to classify target diseases   {\citep{singh2020classification,gao2021covid,ye2022robust,xie2018knowledge,ouyang2020dual,ardila2019end,qiblawey2021detection}}. In these methods, the presence or absence of a specific disease in each CT slice is annotated {\citep{singh2020classification,gao2021covid,ye2022robust,xie2018knowledge}}, or spatial information of the lesion is used within each CT slice   {\citep{ouyang2020dual,ardila2019end,qiblawey2021detection}}. However, to preserve privacy of medical data, it becomes costly to collect disease data and perform expert annotation {\citep{willemink2020preparing,nakao2021unsupervised}}. Consequently, data scarcity of a target disease for diagnosis or insufficient annotation information is common in clinical practice. Hence, conventional SL methods may increase the false positive rate of anomaly detection because training may not provide the expected performance {\citep{kim2021one}}. Moreover, experimental and technical verifications have not been conducted on the prevention of erroneous detection by correctly diagnosing unseen/untrained disease images {\citep{zhang2020viral,schlegl2019f}}. This limited research hinders the use of LDCT to correctly detect any abnormal disease during initial screening of patients, especially regarding the use of SL methods for early diagnosis of new infectious diseases, such as new variants of the coronavirus disease (COVID-19), for which insufficient training data may be available.

\begin{figure*}[t]
\centerline{\includegraphics[width=0.8\textwidth]{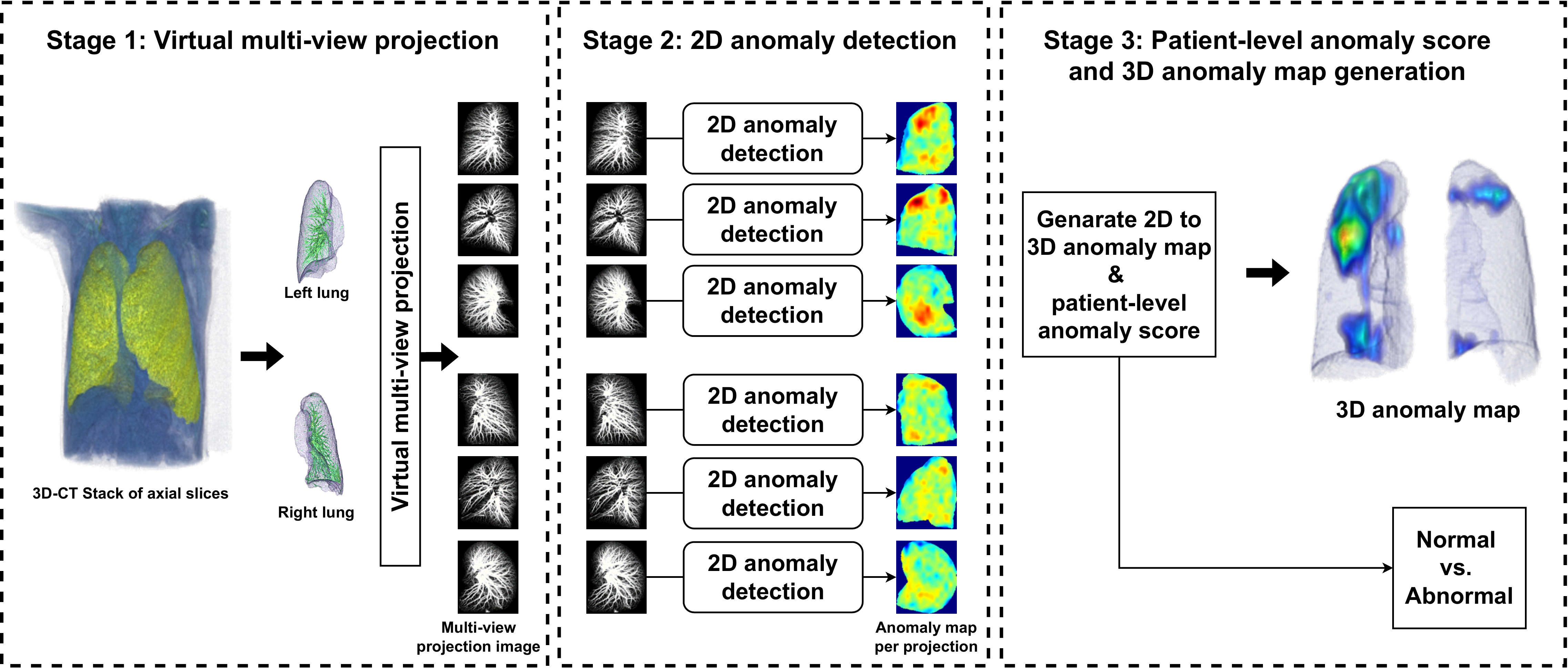}} 
\caption{Overview of proposed VMPR-UAD applied to 3D stack of chest CT slices.}
\label{fig_overview}
\end{figure*}

\subsection{Unsupervised learning applied to LDCT}  
\,\,\,\,\,\,\,\, Unsupervised-learning-based anomaly detection (UAD) has been proposed to improve the diagnostic performance and convenience. Unlike SL methods, a UAD method allows the underlying network to learn by using only data from healthy subjects and then distinguish unseen abnormal conditions. Therefore, anomaly detection can be achieved even when data from a disease are scarce compared with normal cases or when annotation of disease samples is expensive. However, existing UAD methods cannot provide high performance in the detection and localization of lung diseases from LDCT scans. In fact, anomaly detection in medical images should be specialized for detecting fine and specific anomalies; however, existing methods lack such specialization.
 
Most deep-learning-based anomaly detection methods reconstruct a virtual normal patient image from an input image by using a generative adversarial network \citep{goodfellow2014generative} or a (variational) autoencoder \citep{kingma2013auto} to then localize the disease by taking the difference between the input image and synthesized image {\citep{bhatt2021unsupervised,schlegl2019f,nakao2021unsupervised,simarro2020unsupervised,van2021anomaly,dey2021asc}}. However, reconstructing complete images may lead to miss microlesions and does not guarantee accurate localization. Consequently, methods that reconstruct complete images provide a limited visualization of microlesions in specific areas and cannot guarantee accurate localization. PatchCore \citep{roth2021towards} is an accurate anomaly detection method; however, it has not been validated for medical imaging. PatchCore compares the similarity between target and normal images per relative spatial position by selecting and comparing local patches for latent features. Thus, it may be suitable for local lesion detection in medical images. In this study, we validated the medical efficacy of PatchCore by modifying it into a 3D image-based medical anomaly detector for LDCT, expanding two-dimensional (2D) image-based anomaly detection to three dimensions.

\subsection{Contributions}  

\,\,\,\,\,\,\,\, We propose a UAD method for LDCT images called virtual multi-view projection and reconstruction for UAD (VMPR-UAD), whose main characteristics are described below and code is available at \url{https://github.com/kskim-phd/VMPR-UAD}.

\subsubsection{Virtual multi-view projection in UAD} 

\,\,\,\,\,\,\,\, As shown in Fig. 1, conventional methods for LDCT/CT-based computer-aided diagnosis are trained for prediction on $N$ axial-view 2D slices (i.e., typical lung CT slices) as their input. However, each slice does not reflect the 3D structure of lungs because of the difficulty to locate a slice along the $z$ axis only from image information. 
Thus, diagnosis methods must individually identify and store the $N$ slices to describe the 3D spatial structure of lungs. Consequently, such methods may require massive data for diagnosing a lesion during training, thus degrading the diagnostic performance. The proposed virtual multi-view projection solves this problem by projecting the 3D lung spatial structure from different views to provide $T$ (e.g., $T=6$) input images for network training. As each projected image contains part of the lung structure, a method trained with these images may better recognize the complete 3D spatial structure of lungs, further improving the detection performance of specific abnormal areas in the lungs. In addition, the proposed virtual multi-view projection can reduce the number, $T$, of CT images (generally 100 or more) to $N$ ($N<T$). In fact, as each projected image corresponds to a 3D lung voxel at a unique viewing angle, $T$ images can contain sufficient local features (e.g., shading information for distinguishing a disease from a normal area) from various viewpoints and global features (e.g., location information for stable training). By reducing the number of required images, projection can reduce the training and testing costs and allows the diagnosis method to improve target localization.

A similar method uses multiangle views of segmented 3D lung regions rather than conventional CT slices \cite{wu2020deep}; however, it does not use projection. Thus, that method cannot process non-central lung lesions unlike our proposal. Moreover, the proposed projection combines truncation-based maximum intensity projection (MIP) with 3D lung segmentation in a unique method, whose superiority is demonstrated by comparing it with other configurations.

\subsubsection{Validation of high-performance anomaly detection on 3D LDCT} 
\,\,\,\,\,\,\,\, To our knowledge, we presented for the first time an AI-based method to detect anomalies in LDCT. Specifically, we introduce a deep learning and multi-projection-based UAD method for 3D LDCT images and verify its high performance on clinical data. The proposed method, which does not require images containing disease information for training, has a higher anomaly detection performance than an SL method, which requires disease data for training. Thus, our method is highly effective and convenient, especially for detecting a previously unseen anomaly. 

We adopt a distributed method for anomaly detection, in which various anomaly detection networks are trained to perform separate inference on different types of projected images. Then, a patient-level anomaly score is obtained from each network, and the result is synthesized through postprocessing to calculate the overall patient-level anomaly score. As the multiple projected images have globally different characteristics, the proposed distributed method can outperform a conventional centralized method with a single network that performs anomaly detection on images used as multichannel inputs. It is also useful to note that the proposed distributed UAD method is designed for any existing 2D image-based UAD technology to be used. In this study, we verified the validity of the proposed UAD technique by using PatchCore \citep{roth2021towards} as it has a lower computational complexity of learning (i.e., no need for network parameter training) and a more intuitive anomaly localization process (i.e., direct comparison between pixels through nearest neighbor search) than other related modern UADs \citep{tian2021constrained,kim2021one}. However, our technology is designed to be generalized so that any future UAD technology with better performance than PatchCore can be used, allowing our multi-projection-based UAD technology to be further advanced in the future.

\subsubsection{3D anomaly localization via 2D-to-3D map reconstruction} 
\,\,\,\,\,\,\,\, We also propose a method for precisely visualizing and localizing lesions in 3D LDCT scans. Presumably, this is the first method that only requires data from healthy subjects for training and then localizes 3D abnormal regions. Therefore, the proposed method can reduce the burden to specialists who perform annotation by automatically segmenting the disease area through training with healthy subjects' data regardless of the scarcity of disease data.

To optimize the localization performance, we propose a novel 3D reconstruction of 2D anomaly maps, which are multi-network outputs, obtained from the distributed anomaly detection. Reconstruction achieves a high 3D lesion localization performance by maximizing the correlation between the 2D maps, which are registered onto a 3D map that is reconstructed spatially by applying reverse projection.

\section{Methods}
\,\,\,\,\,\,\,\, The proposed VMPR-UAD method is illustrated in Fig. \ref{fig_overview}. It mainly consists of the following three stages. 1) Acquire multiple 2D images by performing virtual projection from a 3D stack of axial CT slices of target patient. 2) Score anomaly and generate 2D anomaly map in each projected image. 3) Generate 3D anomaly map by combining 2D anomaly detection of each projection image and derive the patient-level anomaly score. The stages for VMPR-UAD are detailed below.

\begin{figure*}[t]
\centerline{\includegraphics[width=0.9\textwidth]{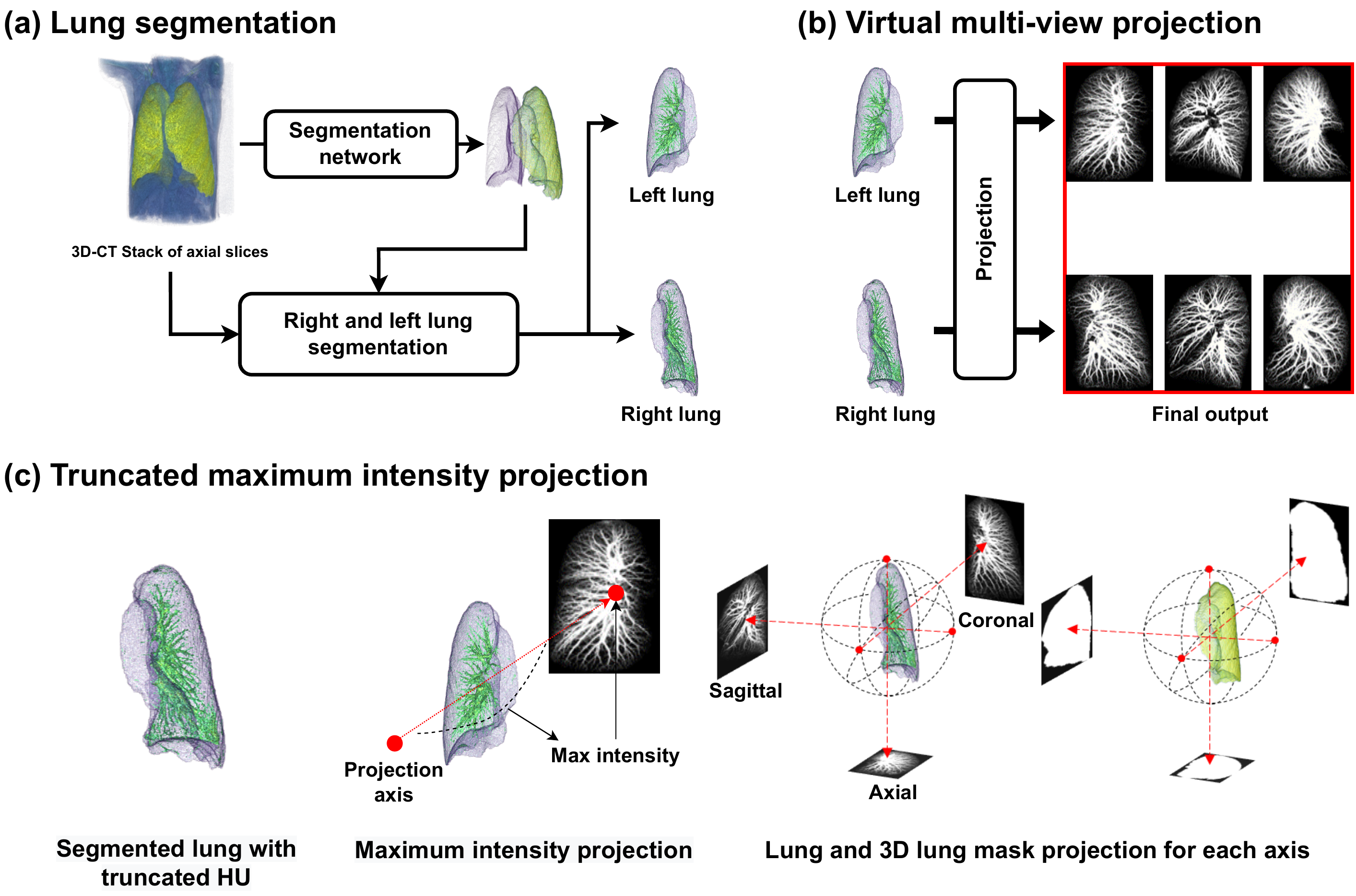}}
\caption{Stage 1 of proposed VMPR-UAD (Virtual multi-view projection). (a) Voxel groups of left and right lungs are divided via lung segmentation on the original 3D stack of axial CT slices. (b) Intensity values are truncated from -800 to 0 Hounsfield units on the 3D voxels of segmented left and right lungs. (c) Multi-view MIP (on axial, coronal, and sagittal planes) is applied to each 3D truncated voxel of the left and right lungs.}
\label{fig_projection}
\end{figure*}

\subsection{Stage 1: Virtual multi-view projection}
\label{stage1-VMPR-UAD}

\,\,\,\,\,\,\,\, Given a 3D stack of axial CT slices, we obtain multiple 2D images by applying multi-view projection to the left and right lung regions on that stack, as specified by the following three steps.  

\subsubsection{Lung segmentation} 
\,\,\,\,\,\,\,\, Before projection, we segment and separate the left and right lungs in the 3D stack of axial CT slices. For segmentation, we use a 3D U-Net \citep{hofmanninger2020automatic}, which provides high-performance extraction of the left and right lung regions from the 3D stack of axial CT slices used as input. Then, projection is performed separately for the left and right lung regions.

\subsubsection{MIP under truncated normalization}
\,\,\,\,\,\,\,\, For projecting the segmented left and right 3D lung regions, we propose MIP under truncated normalization, which provided the highest anomaly detection performance in our experiments. The proposed MIP truncates only the intensity values ranging from -800 to 0 HU (Hounsfield units) and applies min-max normalization to the truncated 3D image. Then, MIP is applied to the normalized 3D image on a given plane. The proposed MIP under truncated normalization retains only the most shaded pixels within the intensity region of interest on a given projection plane, allowing to observe in detail suspicious abnormal structures of the 3D lung stack on the projection plane.

MIP has been widely used for various types of medical data of structures without bones in applications such anomaly detection on angiographic images \citep{venema2001ct,byun2003evaluation}. However, MIP has not been applied to lung CT slices because its use in CT without segmenting the lung regions provides a low sensitivity for anomaly detection owing to the presence of bones surrounding the lung. The proposed method can obtain high diagnostic performance by applying MIP because the lung region of CT is segmented, removing bone structures and leaving only blood vessels and disease structures in the lungs. To further increase the diagnostic performance, the original CT image is truncated to the (-800,0) HU region, where disease shadows have been reported to be prominent \citep{win2014areas,sul2019volumetric,grassi2021evolution}. We verify the superiority of the proposed MIP by comparing the experimental performance of different projection methods.

\subsubsection{Virtual multi-view projection}
\,\,\,\,\,\,\,\, We obtain multiple projected images by applying MIP under truncated normalization on three typical (i.e., sagittal, coronal, and axial) planes and the left and right 3D lung regions, as illustrated in Fig. \ref{fig_projection}(b). Multi-view projection from various angles allows to detect an abnormal area from different angles that would not be found otherwise owing to occlusion by blood vessels at a fixed angle. 

We obtain six 2D projected images, $x_{(r,s)}$, $x_{(r,c)}$, $x_{(r,a)}$, $x_{(l,s)}$, $x_{(l,c)}$, and $x_{(l,a)}$ for the left ($l$) and right ($r$) 3D lung regions on the sagittal ($s$), coronal ($c$), and axial ($a$) planes from a 3D stack of CT slices per patient. Note that this configuration can be changed by performing projections from other angles.
 
\begin{figure*}[t]
\centerline{\includegraphics[width=0.9\textwidth]{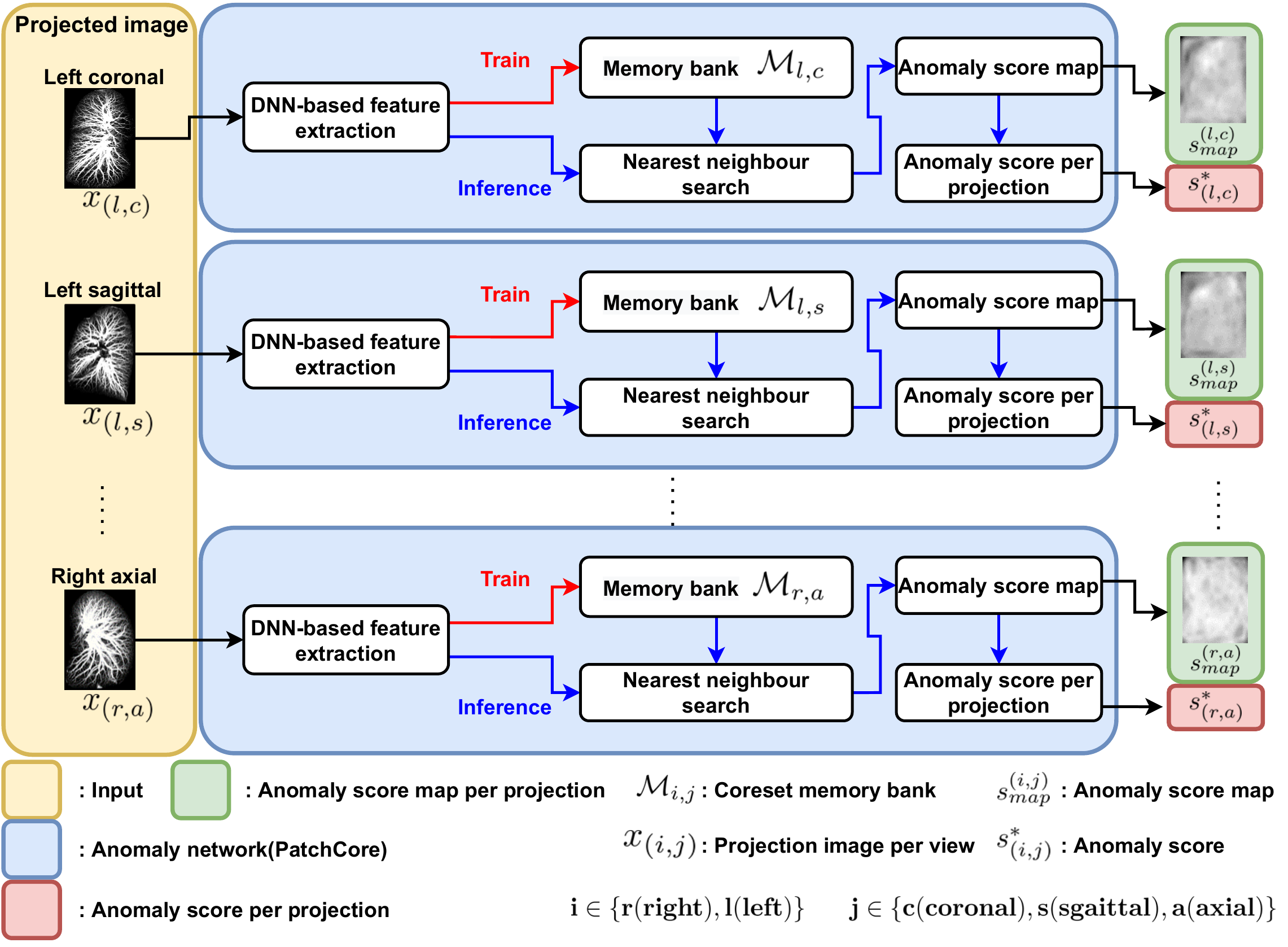}} 
\caption{Stage 2 of proposed VMPR-UAD. Distributed anomaly detection based on multi-view projection images (DNN, deep neural network).}
\label{fig:patchcore_p}
\end{figure*}

\subsection{Stage 2: Distributed anomaly detection for projection images}
\,\,\,\,\,\,\,\, In stage 2, we perform anomaly detection on each of the six projected images, $\{x_{p}\}_{p=(p_1,p_2), p_1 \in \{r,l\},  p_2 \in \{s,c,a\}}$, obtained from stage 1. The six projection images are obtained from CT scans of healthy subjects through stage 1, and individual anomaly detection is applied to the data of each projection.
 
To perform anomaly detection per projected image, we use PatchCore, which can be converted to other networks. PatchCore {\citep{roth2021towards}} consists of 1) local patch feature extraction of original images, in which training data are aggregated into memory bank $\mathcal{M}$; 2) coreset-sampling local patch features in the bank for reducing inference complexity; and 3) anomaly detection and localization through nearest neighbor search between the bank and test data. 

Given nominal (coreset-sampled) patch-feature memory bank $\mathcal{M}$, PatchCore obtains the original target image, $x^{test}$, as input and provides the 2D anomaly map, $s_{map}$, by finding the pixelwise minimum distance between the feature vectors of the target image and those of normal samples at location $ij$ as follows:
\begin{align}\nonumber 
      m^{test,*}_{ij}, m^{*}_{ij} &= \underset{m^{test}_{ij} \in \mathcal{P}_{ij}(x^{test})}{\argmax} \underset{m \in \mathcal{M}}{\argmin} \left\| m^{test}_{ij} - m \right\|, \\\label{2dmap}
      s_{map}(i,j) &=  \left\| m^{test,*}_{ij} - m^{*}_{ij}\right\|, \\\label{ab_score}
      s^* &=  \underset{i,j}{\max} \,\, s_{map}(i,j),
\end{align}
where $\mathcal{P}_{ij}(x^{test})$ denotes the feature vector of target image $x^{test}$ at location $ij$. The patient-level anomaly score is given by $s^*$ in Eq. \eqref{ab_score} by finding the pixelwise maximum in $s_{map}$, corresponding to the most abnormal part.

Anomaly detection per projected image is performed using PatchCore by extending the derivations above as follows:
\begin{align}\nonumber 
      m^{test,*}_{ij,p}, m^{*}_{ij,p} &= \underset{m^{test}_{ij} \in \mathcal{P}_{ij}(x^{test}_{p})}{\argmax} \underset{m \in \mathcal{M}_{p}}{\argmin} \left\| m^{test}_{ij,p} - m \right\|, \\\label{2dmap_p}
      s^p_{map}(i,j) &=  \left\| m^{test,*}_{ij,p} - m^{*}_{ij,p}\right\|, \\\label{ab_score_p}
      s^*_p &=  \underset{i,j}{\max} \,\, s^p_{map}(i,j),
\end{align}
where $p=(p_1,p_2)$ ($p_1 \in \{r,l\},  p_2 \in \{s,c,a\}$) denotes the index representing one of the six projection types and $\mathcal{M}_{p}=\{m^{train,i}_{p}\}_{i=1}^C$ denotes the memory bank of $C$ coreset-sampled patch feature vectors, which is extracted from $p$-type projected training images $\{x^{train,i}_{p}\}_{i=1}^K$ of $K$ normal patients. Therefore, the 2D anomaly map and the anomaly score for target projected image $x^{test}_{p}$ can be obtained as $s^p_{map} \in \mathbb{R}^{h \times w}$ and $s^*_p \in \mathbb{R}$ in Eqs. \eqref{2dmap_p} and \eqref{ab_score_p}, respectively. The process of stage 2 is illustrated in Fig. \ref{fig:patchcore_p}.

\begin{figure*}[t]
\centerline{\includegraphics[width=0.93\textwidth]{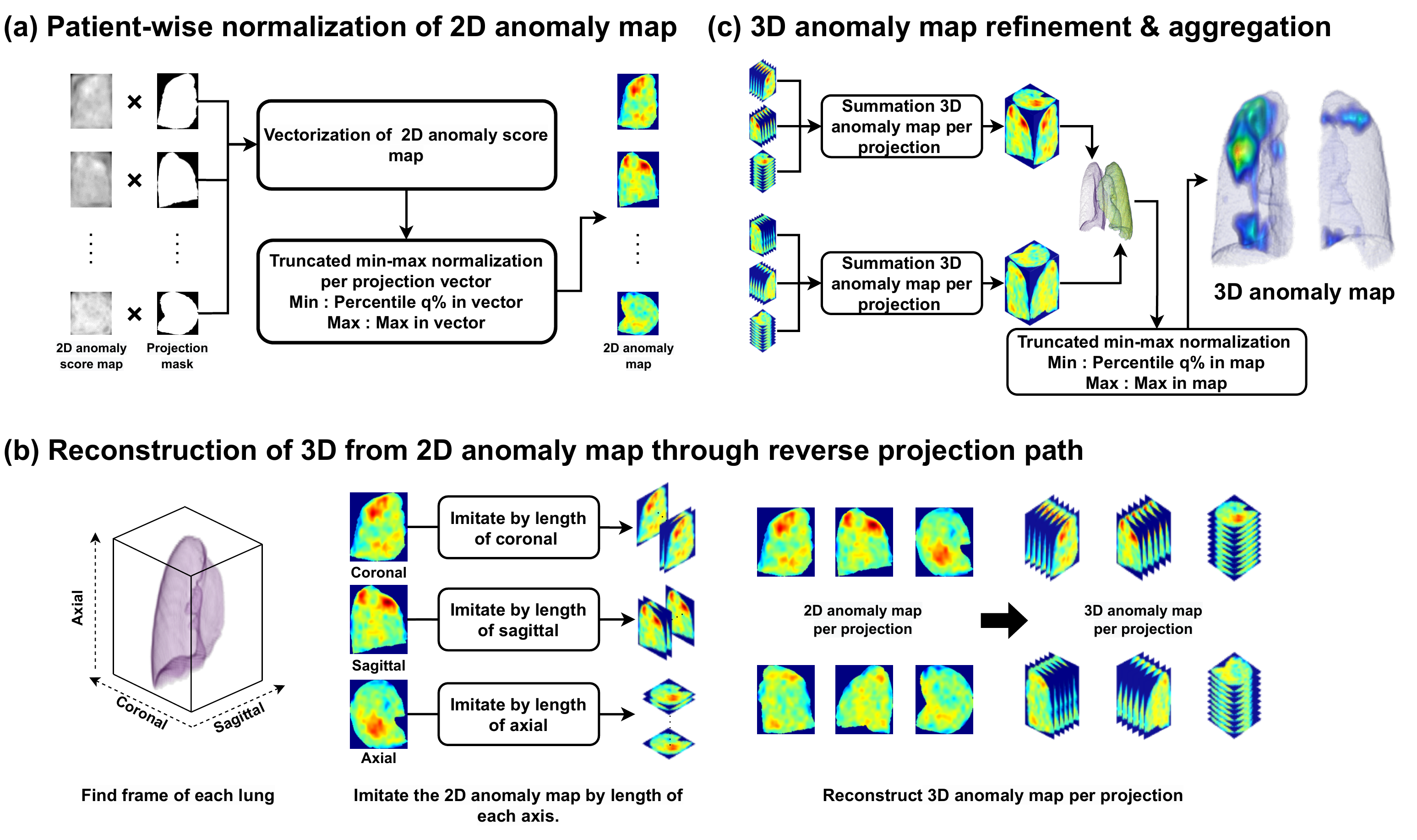}} 
\caption{Stage 3 of proposed VMPR-UAD. Proposed 3D anomaly map generation.}
\label{fig:3dcam}
\end{figure*}

\subsection{Stage 3: 2D-to-3D anomaly map reconstruction and patient-level anomaly detection}
\,\,\,\,\,\,\,\, In stage 3, we obtain the 3D anomaly map (i.e., $v^l_{3d\textup{-}map}$ and $v^r_{3d\textup{-}map}$ for the left and right lungs, respectively) through 3D comprehensive reconstruction of the projected 2D anomaly maps, $\{s^{p}_{map}\}_{p=(p_1,p_2), p_1 \in \{r,l\},  p_2 \in \{s,c,a\}}$, through the corresponding reverse projection. In addition, we derive the patient-level anomaly score quantifying the anomaly degree in a patient. These processes can be divided into three steps illustrated in Fig. \ref{fig:3dcam} and formulated as follows:

\begin{align}\label{st3:firststep} 
    r^p_{map} &= \mathcal{T}_{pmm,q}(z^p_{2d\textup{-}mask} \odot s^p_{map}), \\\label{st3:2thstep}
    u^p_{3d\textup{-}map} &= \mathcal{T}_{recon}(r^p_{map};p),\\\label{st3:3thstep_1} 
    u^k_{3d\textup{-}map} &= z^k_{3d\textup{-}mask} \odot \Big[ \sum_{i \in \{s,c,a\}} u^{(k,i)}_{3d\textup{-}map} \Big] \,\,\,\,\,\,\,\,\, (k\in\{r,l\}), \\\label{st3:3thstep_2} 
    v_{3d\textup{-}map} &= \mathcal{T}_{pmm,q}\Big( u^r_{3d\textup{-}map}, u^l_{3d\textup{-}map} \Big). 
\end{align}

\subsubsection{2D-to-3D anomaly map reconstruction} 
\,\,\,\,\,\,\,\, As shown in Eq. \eqref{st3:firststep} and Fig. \ref{fig:3dcam}(a), the region of interest is extracted in anomaly map $s^p_{map}$ of each projected image. For extraction, we propose percentile-based min-max normalization $\mathcal{T}_{pmm,q}(z^p_{2d\textup{-}mask} \odot s^p_{map})$, which efficiently removes low-probability anomaly positions. Specifically, the proposed normalization removes all pixel values below the value corresponding to the $q\%$ percentile ($q=50$ in this study) from the distribution of pixel values in input image $z^p_{2d\textup{-}mask} \odot s^p_{map}$. This input image corresponds to subset $s^p_{map}$ of anomaly map inside the lungs, and $z^p_{2d\textup{-}mask}$ corresponds to a binary mask of $s^p_{map}$ obtained by applying the same $p$-type projection to the 3D binary mask lung region. The proposed normalization emphasizes the anomaly candidate region, thereby improving anomaly detection and localization. 

Then, virtual reconstruction $u^{p}_{3d\textup{-}map}=\mathcal{T}_{recon}(r^p_{map};p)$ of each 2D anomaly map $r^p_{map}$ into a 3D representation is performed by repeatedly arranging every 2D anomaly map along direction $p$ of projection, as shown in Eq. \eqref{st3:2thstep} and Fig. \ref{fig:3dcam}(b). To each of the obtained 3D images, we apply the proposed normalization in three dimensions to highlight the 3D region of the anomaly candidate. 

Finally, as shown in Eqs. \eqref{st3:3thstep_1} and \eqref{st3:3thstep_2} and Fig. \ref{fig:3dcam}(c), three 3D anomaly maps are averaged for each of the left and right lung regions (i.e., $\sum_{i \in \{s,c,a\}} u^{(c,i)}_{3d\textup{-}map}$ for $c \in \{l,r\}$) and filtered/normalized by using the corresponding 3D region mask ($u^l_{3d\textup{-}map}$ and $u^r_{3d\textup{-}map}$ for the left and right lungs, respectively) to obtain the final 3D anomaly map, $v_{3d\textup{-}map}$, which covers the left and right lung regions by extracting only high-probability anomaly positions via the proposed percentile-based min-max normalization. 

\subsubsection{Patient-level anomaly score} 
\,\,\,\,\,\,\,\, We obtain the patient-level anomaly score, $s^*$, that quantifies the anomaly degree from 0 to 1 (a larger value indicates a severer anomaly) by averaging anomaly scores $s^*_p$ of the 2D projections across the six projection types, $p=(p_1,p_2)$ with $p_1 \in \{r,l\},  p_2 \in \{s,c,a\}$. In Section \ref{comp_projnum}, we show that the proposed anomaly score obtained from the projection-level average outperforms the anomaly score of individual projections.

\section{Experimental results}

\subsection{Evaluation metrics}
\,\,\,\,\,\,\,\, To evaluate the anomaly detection performance of the proposed VMPR-UAD (i.e., binary classification of normal or abnormal due to lung disease of patients), we used the area under the curve (AUC) of the receiver operating characteristic (ROC) curve, accuracy, sensitivity, precision, and F1 score. In the ROC curve, we selected the the decision threshold closest to the point where the false and true positive rates were 0 and 1, respectively, and calculated the true positive ($TP$), true negative ($TN$), false positive ($FP$), and false negative ($FN$) rates based on that threshold. Then, the accuracy, sensitivity, precision, and F1 scores were calculated as follows:
\begin{align}\nonumber  
    Accuracy &= \frac{(TN+TP)}{(TN+TP+FN+FP)},  \\\nonumber 
    Sensitivity &= \frac{(TP)}{(TP+FN)}, \\\nonumber 
    Precision &= \frac{(TP)}{(TP+FP)}, \\\nonumber 
    F1\, score &= 2 \, \frac{Precision \times Sensitivity}{Precision+Sensitivity}.
\end{align}
\subsection{Data preparation}
\,\,\,\,\,\,\,\, For internal validation, we collected 7560 3D stacks of DICOM (Digital Imaging and Communications in Medicine) LDCT slices from Samsung Medical Center (SMC), Seoul, Republic of Korea, as specified in Table \ref{tab:datasets_info}. This validation was retrospective and performed in accordance with the Declaration of Helsinki and relevant scientific guidelines. The study protocols were approved by the institutional review board of our institution SMC (approval number: 2021-11-076). The requirement for informed consent was waived owing to the retrospective nature of this study. The 7560 3D stacks were acquired from 6660 healthy, 700 pneumonia, and 200 tuberculosis cases. We also performed an external validation using the publicly available lung cancer dataset of the medical segmentation decathlon (MSD) \citep{antonelli2021medical}. 

To train and validate the proposed VMPR-UAD, we adopted Monte Carlo validation. Specifically, 4000 out of the 6660 stacks from healthy subjects were randomly selected for network training (i.e., collecting data in each coreset memory bank $\mathcal{M}_p$ per projection type $p$). For testing, we evaluated the performance of binary anomaly classification for a single and two diseases by adjusting the number of evaluation stacks of normal and abnormal patients in a one-to-one ratio. Specifically, for pneumonia detection, 700 of the remaining 2660 stacks from healthy subjects were randomly selected along with the 700 stacks showing pneumonia. For tuberculosis detection, 200 of the 2660 stacks from healthy subjects were randomly selected along with the 200 stacks showing tuberculosis. For multiple-disease detection, 900 of the 2660 stacks from healthy subjects were randomly selected along with the 700 and 200 stacks showing pneumonia and tuberculosis, respectively. For the external validation, 95 of the 2660 stacks from healthy subjects were randomly selected along with the 95 stacks showing lung cancer. We repeated each experiment five times with fivefold Monte Carlo validation.

\begin{table}[hbt!]
	\vskip -3pt 
	\caption{\footnotesize {Datasets used in this study. Number of samples in each dataset (left) and SMC dataset specifications (right), where the tube voltage of 120 kVp indicates LDCT.}}
	\footnotesize
	\centering
	{
	\subtable{
		\resizebox{0.47\linewidth}{!}{
			\begin{tabular}{ccccc}
				\toprule
			    Dataset & Label &\:Total\:& \:Train\: & \:Inference\: \\
				\midrule
			
			    \multirow{3}{*}{SMC (Internal dataset)}
			    &Healthy  &6660& {4000} &900\\
				&Pneumonia &700& {--} & 700 \\
				&Tuberculosis  &200&{--} &	200  \\
				
				\midrule
				MSD \citep{antonelli2021medical}
				&Cancer &95 &{--}&95 
				 \\
				\bottomrule
			\end{tabular}
			
		} 
      \label{tab:datasets1}
		}
	\subtable{	
		\resizebox{0.5\linewidth}{!}{
			\begin{tabular}{cc}

		      	\toprule
			    \small{Parameter} & \small{Value} \\
				\midrule
			   \small{Tube current time product}&  \small{16--139 mAs}\\
			   \small{Tube voltage}&  \small{120 kVp}\\
			   \small{View}& \small{Axial plane}\\
			   \small{Slice size} & \small{512 $\times$ 512 pixels}\\
			   \small{Number of slices}&  \small{135\:--\:390}\\
			   \small{Pixel spacing}& \:\: \small{0.484 $\times$ 0.484${mm}^{2}$\:--\:0.887 $\times$ 0.887${mm}^{2}$}\:\:\\
			   \small{Slice thickness}& \small{0.625, 1, 1.25, 5}\\
				\bottomrule

			\end{tabular}
		} 
	\label{tab:datasets2}
	}
	}
	\label{tab:datasets_info}
	\vspace{-0.2cm}
\end{table}

\begin{figure}[t]
	\vskip -5pt
	\centering
	\subfigure{\includegraphics[width=0.85\linewidth]{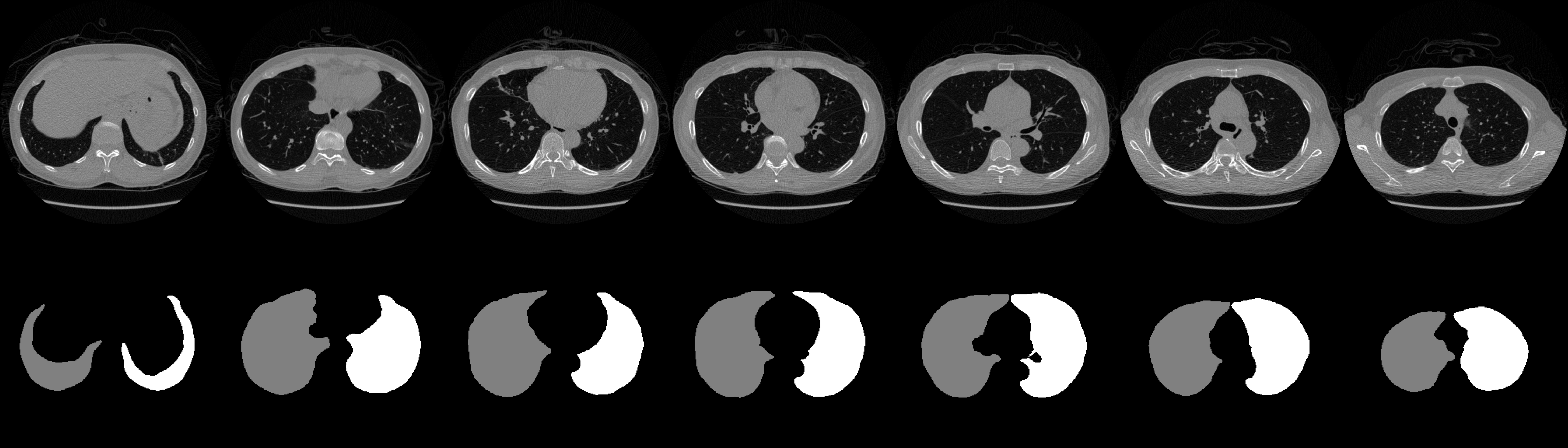}}
	\vskip -4pt

	\subfigure{\includegraphics[width=0.85\linewidth]{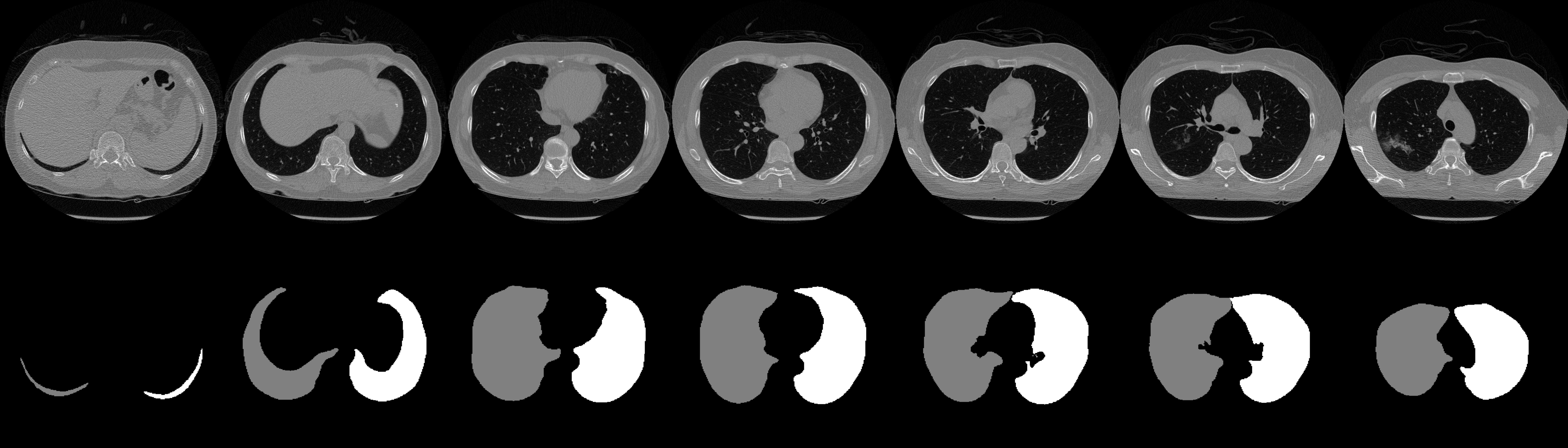}}
	\vskip -4pt

	\subfigure{\includegraphics[width=0.85\linewidth]{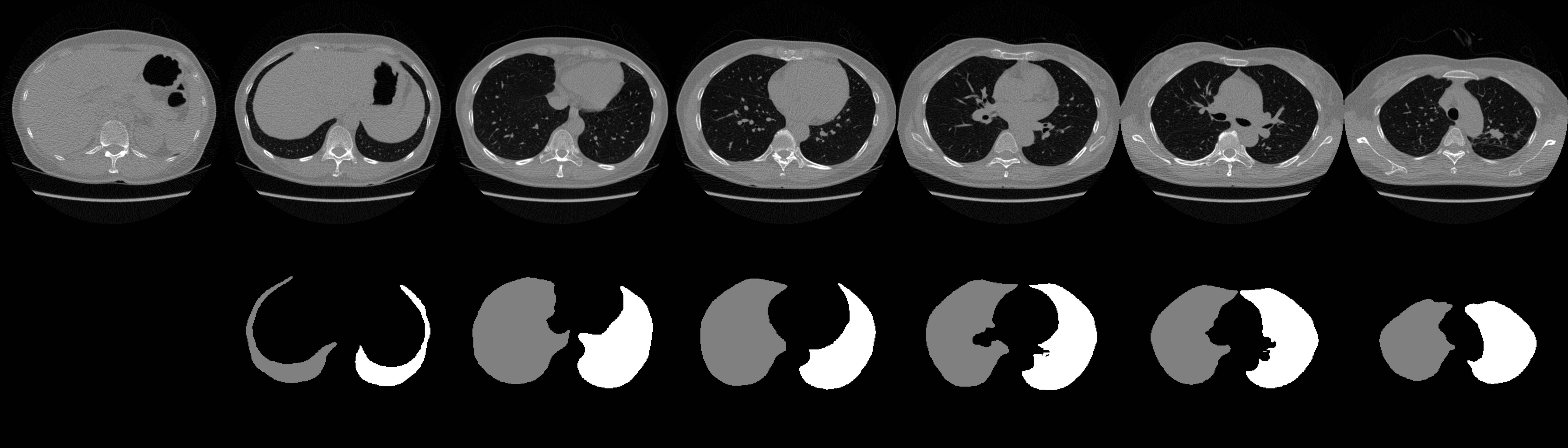}}
	\vspace{-0.2cm}
	\caption{\footnotesize Lung masks predicted by segmentation network (U-net-R231) for normal (top), pneumonia (middle), and tuberculosis (bottom) cases.}
	\label{fig:ref_seg_mask}
\end{figure}

\subsection{Segmentation} 

\begin{wraptable}{r}{6cm}
	\vskip -15pt 
	\caption{\footnotesize {Statistical results of 3D lung segmentation}}
	\centering
	{
		\resizebox{\linewidth}{!}{

			\begin{tabular}{cccc}
				\toprule
			     &{Data} & {IoU} & {Dice}\\
				\midrule
			    \multirow{3}{*}{{Reference study}}&
			    {LTRC}& {--}& {$0.99$} \\
			 &{LCTSC}& {--}&{$0.94$}\\
			 &{VESSEL12}& {--}&{$0.98$}\\
				\midrule
			    {Our study} &{COVID-19} & 
				{$0.951$}  &
				{$0.975$}  \\
				\bottomrule
			\end{tabular}
		
		}
	}
	\label{tab:segment_result}
 \vspace{-10pt}
\end{wraptable}

\,\,\,\,\,\,\,\, We used a publicly available pretrained lung segmentation network, U-net-R231 \citep{hofmanninger2020automatic}, to segment lung regions from each 3D stack of CT slices. We used the COVID-19 dataset \citep{ma_jun_2020_3757476} to evaluate the lung segmentation performance. The intersection of the ground truth for the 3D lung region with the U-net-R231 prediction was evaluated using the intersection over union (IoU) and Dice score and averaged across slices and stacks, obtaining the results listed in Table \ref{tab:segment_result}. Compared with the performance evaluation results using the datasets in \citep{hofmanninger2020automatic} (i.e., lung tissue research consortium–--LTRC \citep{karwoski2008processing}, lung CT segmentation challenge 2017–--LCTSC \citep{yang2018autosegmentation}, and vessel segmentation in the lung 2012–--VESSEL12 \citep{rudyanto2014comparing}) also shown in Table \ref{tab:segment_result}, U-net-R231 provides a high Dice score, thereby validating its excellent lung segmentation ability. Fig. \mbox{\ref{fig:ref_seg_mask}} shows results of lung segmentation by U-net-R231 to illustrate its excellent performance.

\begin{table}[hbt!]
	\caption{\footnotesize {Statistical results of proposed VMPR-UAD and SL baseline for anomaly detection. The mean and standard deviation were obtained from fivefold Monte Carlo validation.}}
	\vskip +5pt
	\footnotesize
	\centering
	{
		\resizebox{.8\linewidth}{!}{
			\begin{tabular}{ccccccc}
				\toprule
			     Method & Label & Accuracy & Sensitivity &Specificity& Precision& F1 score \\

				\midrule
					
			    \multirow{3}{*}{SL}
				&Pneumonia  & 
				0.826 $\pm$ 0.09 & 
				0.826 $\pm$ 0.02 & 
				0.826 $\pm$ 0.16 & 
				0.840 $\pm$ 0.12 & 
				0.830 $\pm$ 0.07 \\
				
				&Tuberculosis  & 
				0.793 $\pm$ 0.11 &
				0.764 $\pm$ 0.05 &
				0.822 $\pm$ 0.17 &
				0.825 $\pm$ 0.14 &
				0.791 $\pm$ 0.09 \\
				
				&Both diseases &
		    	0.816 $\pm$ 0.09 & 
				0.809 $\pm$ 0.03 & 
        		0.824 $\pm$ 0.16 &
        		0.835 $\pm$ 0.12 & 
        	    0.820 $\pm$ 0.08 \\
        		
				\midrule

			    \multirow{3}{*}{Ours}
				&Pneumonia  & 
				\textbf{0.907 $\pm$ 0.01} &
				\textbf{0.902 $\pm$ 0.01} &
				\textbf{0.913 $\pm$ 0.01} & 
				\textbf{0.912 $\pm$ 0.01} & 
				\textbf{0.907 $\pm$ 0.01}  \\
				
				&Tuberculosis  & 
				\textbf{0.893 $\pm$ 0.02} &
				\textbf{0.889 $\pm$ 0.01} & 
				\textbf{0.897 $\pm$ 0.03} & 
				\textbf{0.897 $\pm$ 0.03} & 
				\textbf{0.893 $\pm$ 0.02}  \\
				
				&Both diseases &
			    \textbf{0.904 $\pm$ 0.01}& 
				\textbf{0.899 $\pm$ 0.01}&
				\textbf{0.910 $\pm$ 0.01} &
				\textbf{0.909 $\pm$ 0.01} &
				\textbf{0.904 $\pm$ 0.01} \\
				
				\bottomrule
			\end{tabular}
		}
	}
	\label{tab:sl_comp}
	\vspace{-0.2cm}
\end{table}

\begin{figure}[hbt!]
	\vskip -2pt
	\centering
	\subfigure[]{\includegraphics[width=0.38\linewidth]{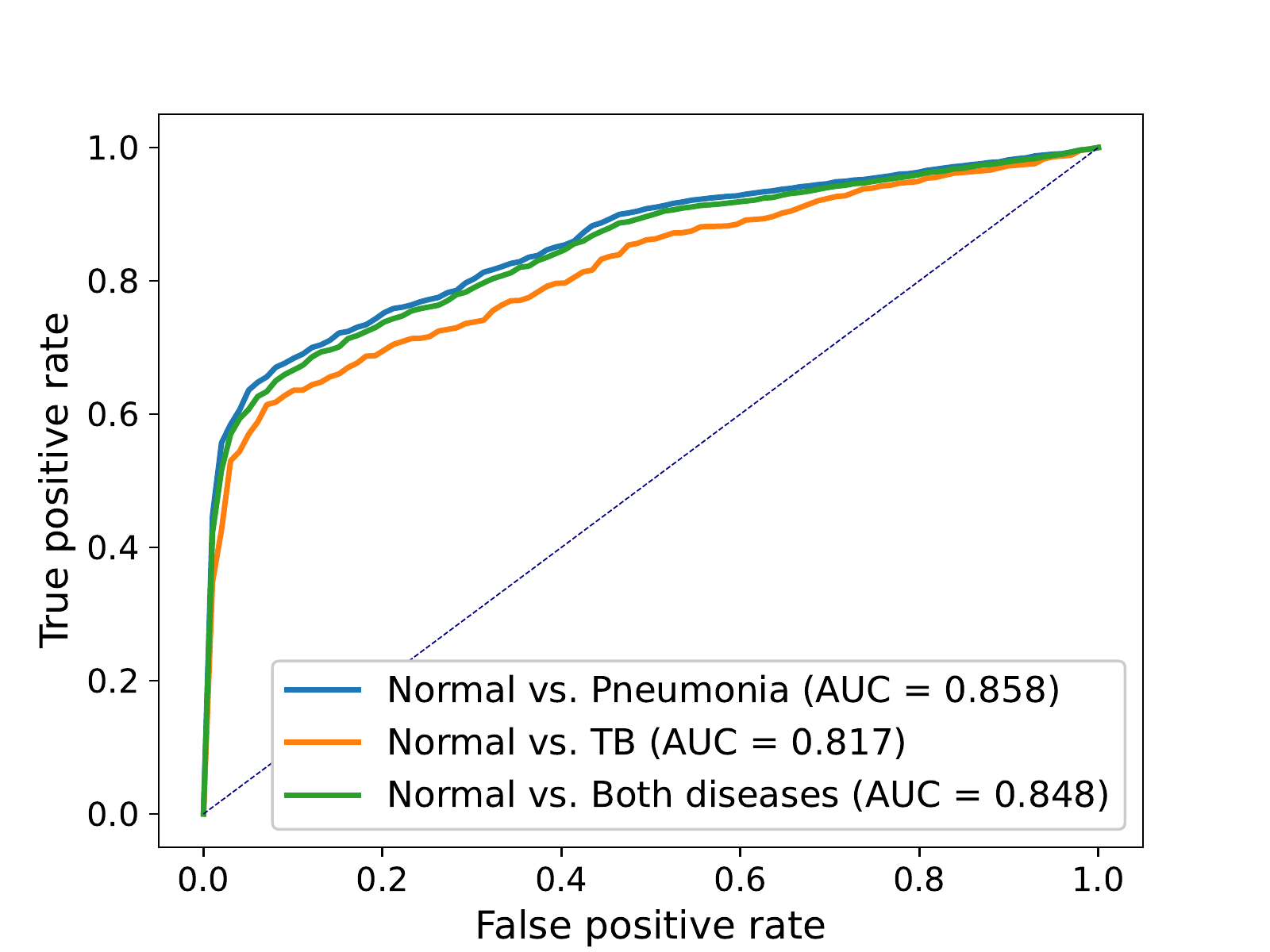}}
	\subfigure[]{\includegraphics[width=0.38\linewidth]{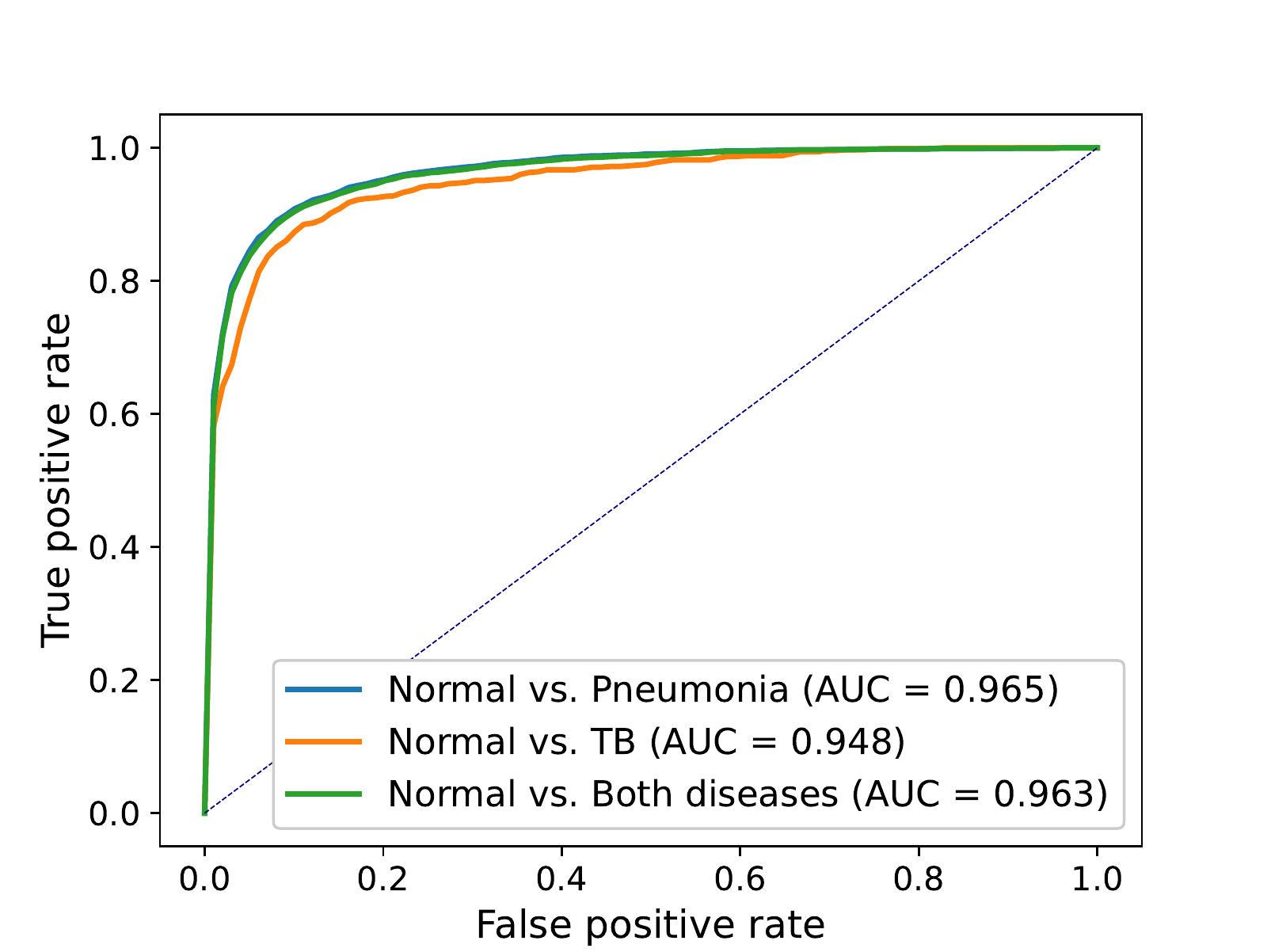}}
	\vspace{-0.2cm}
	\caption{\footnotesize ROC curves of (a) SL baseline and (b) proposed VMPR-UAD. The mean was obtained from fivefold Monte Carlo validation. }
	\vspace{-0.2cm}
	\label{fig:sl_comp}
\end{figure}

\subsection{Internal validation for LDCT data of pneumonia and tuberculosis cases}

\,\,\,\,\,\,\,\, By using the internal clinical dataset from SMC of pneumonia and tuberculosis cases, we experimentally verified the superiority of the proposed VMPR-UAD through various comparative experiments. We also provide quantitative results and 3D abnormal (lesion) localization/visualization results of VMPR-UAD using the internal dataset for pneumonia and tuberculosis in the Supplementary Material to illustrate the superiority of anomaly localization by VMPR-UAD.

\subsubsection{Performance of VMPR-UAD and SL method}
\label{comp_sl}

\,\,\,\,\,\,\,\, The proposed VMPR-UAD only requires data from healthy subjects for network training. To compare the VMPR-UAD with an SL method that requires training data from both healthy subjects and patients showing anomalies, the SL method was trained as follows. We set the number of training samples to the maximum possible, with the ratio of normal to abnormal cases being 1:1, and fivefold Monte Carlo validation was adopted during SL training. We randomly selected 450 normal, 350 pneumonia, and 100 tuberculosis cases for training. For pneumonia detection using the SL method, 350 of the remaining stacks from healthy subjects were randomly selected along with 350 stacks showing pneumonia. For tuberculosis detection, 100 stacks from healthy subjects were randomly selected along with the 100 stacks showing tuberculosis. For detection of both diseases, 450 stacks from healthy subjects were randomly selected along with the 350 and 100 stacks showing pneumonia and tuberculosis, respectively. For the SL network, we adopted a 3D convolutional neural network, 3D-ResNet18 {\citep{hara2018can}}, with the 3D stack of original CT slices as its input. The results of comparing the SL method with the proposed VMPR-UAD are listed in Table \ref{tab:sl_comp} and shown in Fig. \ref{fig:sl_comp}. The proposed VMPR-UAD shows higher performance in anomaly detection by improving the AUC by more than 10\%, despite only using data from healthy subjects for training. This result confirms the superiority and convenience of our proposal over an SL-based diagnosis, especially when the training data of the target disease are insufficient or difficult to distinguish and annotate.


\begin{table}[hbt!]
	\vskip -3pt 
	\caption{\footnotesize {Statistical results of proposed VMPR-UAD according to number of projections. The mean and standard deviation were obtained from fivefold Monte Carlo validation.}}
	\footnotesize
	\centering
	{
		\resizebox{.95\linewidth}{!}{
			\begin{tabular}{ccccccc}
				\toprule
			    No. projections & Label & Accuracy & Sensitivity &Specificity& Precision& F1 score \\
				\midrule
			    \multirow{3}{*}{Two projections (coronal)}
				&Pneumonia  & 
				0.858 $\pm$ 0.01 &
				0.849 $\pm$ 0.01 &
				0.868 $\pm$ 0.02 & 
				0.865 $\pm$ 0.02 & 
				0.857 $\pm$ 0.01  \\
				
				&Tuberculosis  & 
				0.863 $\pm$ 0.02 &
				0.853 $\pm$ 0.03 & 
				0.873 $\pm$ 0.03 & 
				0.871 $\pm$ 0.03 & 
				0.862 $\pm$ 0.02  \\
				
				&Both diseases &
			    0.859 $\pm$ 0.01& 
				0.852 $\pm$ 0.02&
				0.866 $\pm$ 0.01 &
				0.864 $\pm$ 0.01 &
				0.858 $\pm$ 0.01 \\
				
				\midrule
				\multirow{3}{*}{Two projections (sagittal)}
				&Pneumonia  & 
				0.875 $\pm$ 0.01 &
				0.857 $\pm$ 0.01 &
				0.894 $\pm$ 0.01 & 
				0.889 $\pm$ 0.01 & 
				0.873 $\pm$ 0.01  \\
				
				& Tuberculosis  & 
				0.859 $\pm$ 0.01 &
				0.847 $\pm$ 0.02 & 
				0.871 $\pm$ 0.01 & 
				0.868 $\pm$ 0.01 & 
				0.857 $\pm$ 0.01  \\
				
				&Both diseases &
			    0.873 $\pm$ 0.01& 
				0.858 $\pm$ 0.01&
				0.887 $\pm$ 0.01 &
				0.884 $\pm$ 0.01 &
				0.871 $\pm$ 0.01 \\
				
				\midrule
			    \multirow{3}{*}{Four projections}
				&Pneumonia  & 
				0.896 $\pm$ 0.01 &
				0.891 $\pm$ 0.02 &
				0.901 $\pm$ 0.01 & 
				0.901 $\pm$ 0.01 & 
				0.896 $\pm$ 0.01  \\
				
				& Tuberculosis  & 
				0.881 $\pm$ 0.02 &
				0.877 $\pm$ 0.02 & 
				0.885 $\pm$ 0.02 & 
				0.884 $\pm$ 0.02 & 
				0.881 $\pm$ 0.02  \\
				
				&Both diseases &
			    0.893 $\pm$ 0.01 &
				0.888 $\pm$ 0.01 & 
				0.899 $\pm$ 0.01 & 
				0.898 $\pm$ 0.01 & 
				0.893 $\pm$ 0.01  \\
				
				\midrule
			    \multirow{3}{*}{Six projections (Ours)}
				&Pneumonia  & 
				\textbf{0.907 $\pm$ 0.01} &
				\textbf{0.902 $\pm$ 0.01} &
				\textbf{0.913 $\pm$ 0.01} & 
				\textbf{0.912 $\pm$ 0.01} & 
				\textbf{0.907 $\pm$ 0.01}  \\
				
				&Tuberculosis  & 
				\textbf{0.893 $\pm$ 0.02} &
				\textbf{0.889 $\pm$ 0.01} & 
				\textbf{0.897 $\pm$ 0.03} & 
				\textbf{0.897 $\pm$ 0.03} & 
				\textbf{0.893 $\pm$ 0.02}  \\
				
				&Both diseases &
			    \textbf{0.904 $\pm$ 0.01}& 
				\textbf{0.899 $\pm$ 0.01}&
				\textbf{0.910 $\pm$ 0.01} &
				\textbf{0.909 $\pm$ 0.01} &
				\textbf{0.904 $\pm$ 0.01} \\
				
				\bottomrule
			\end{tabular}
		}
	}
	\label{tab:number_of_projection}
	\vspace{-0.2cm}
\end{table}

\begin{figure*}[hbt!]
	\vskip -5pt
	\centering
	\subfigure[]{\includegraphics[width=0.38\linewidth ]{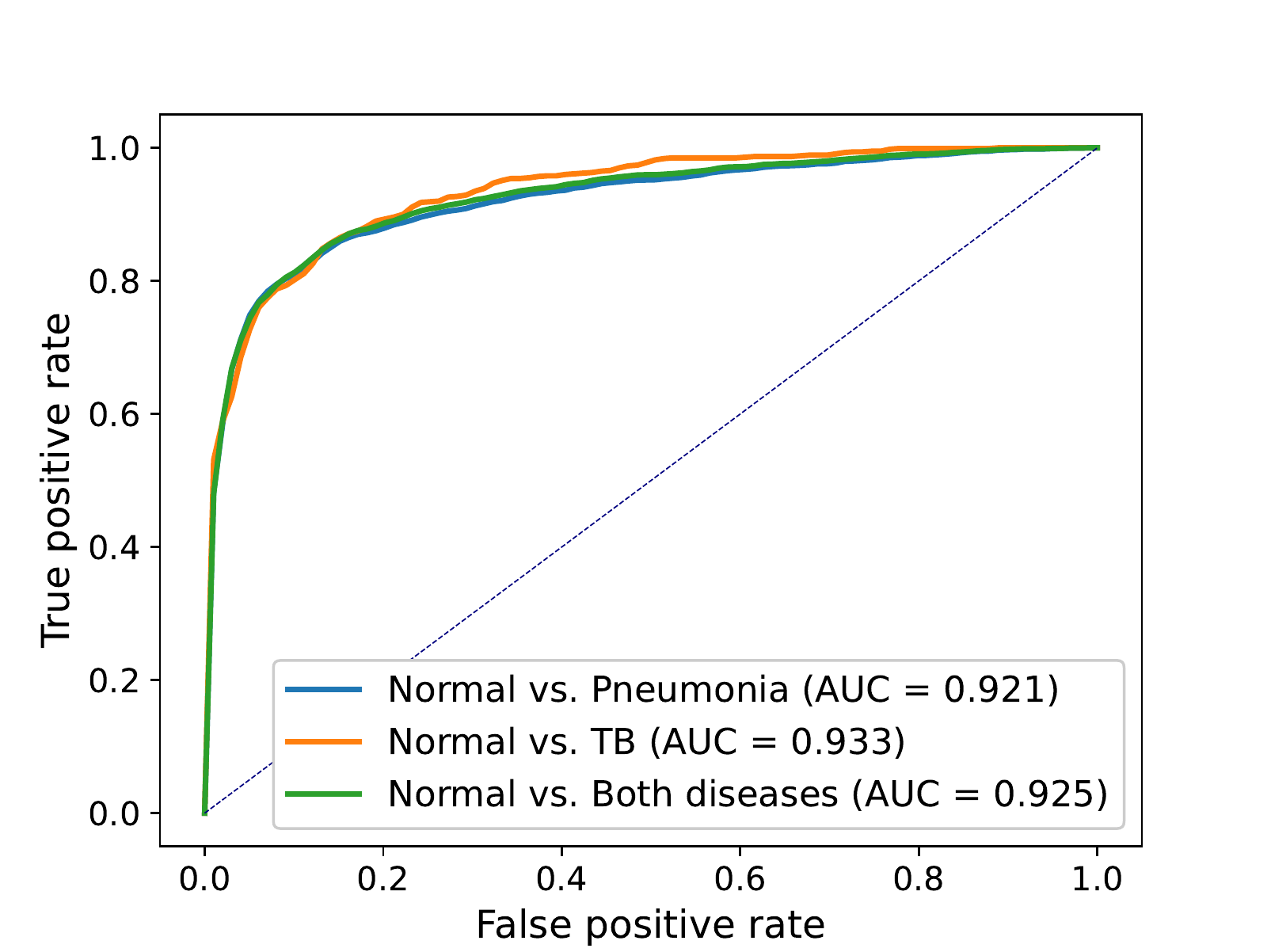}}
	\subfigure[]{\includegraphics[width=0.38\linewidth ]{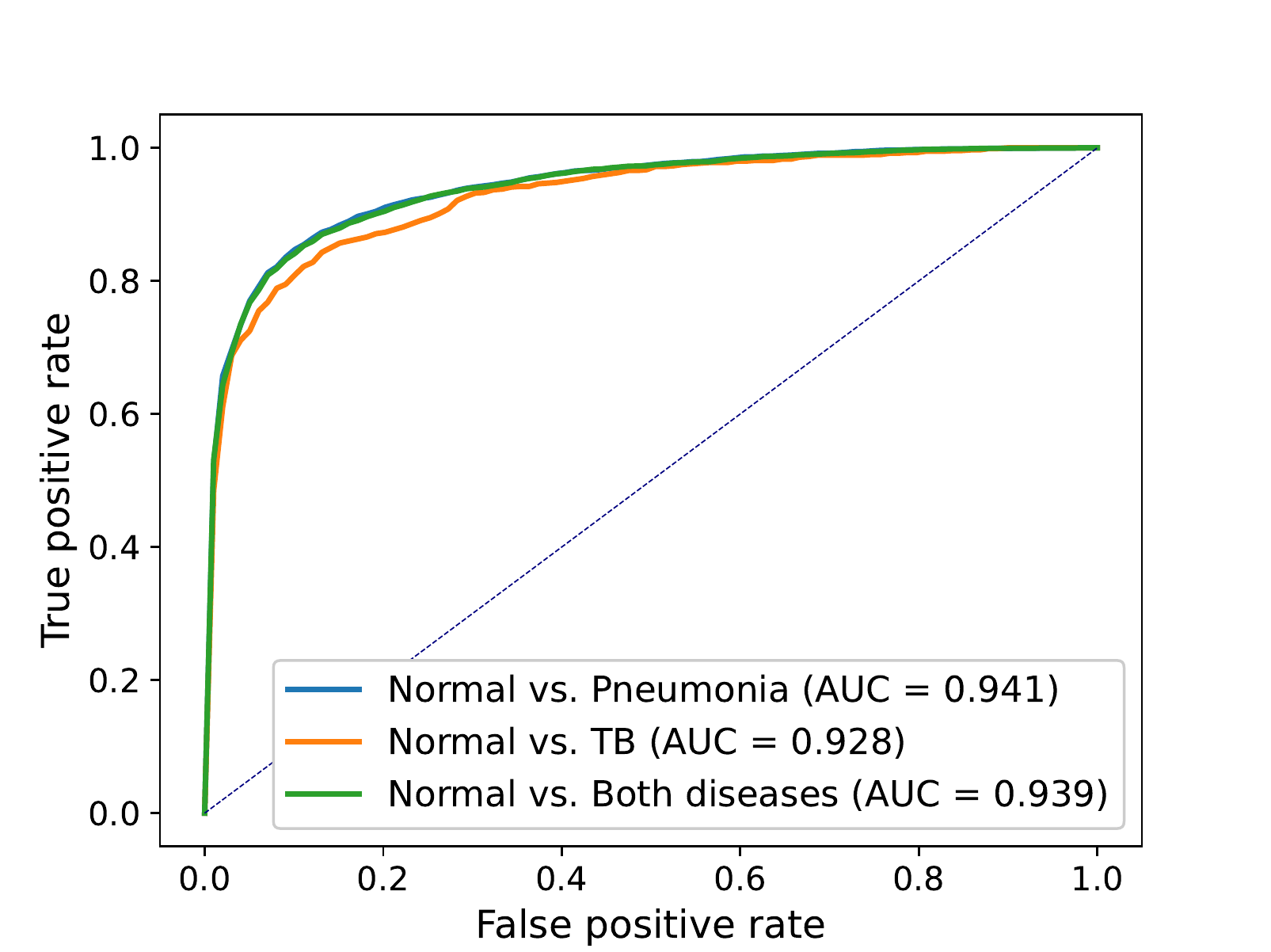}}
	\vskip -25pt
	\subfigure[]{\includegraphics[width=0.38\linewidth ]{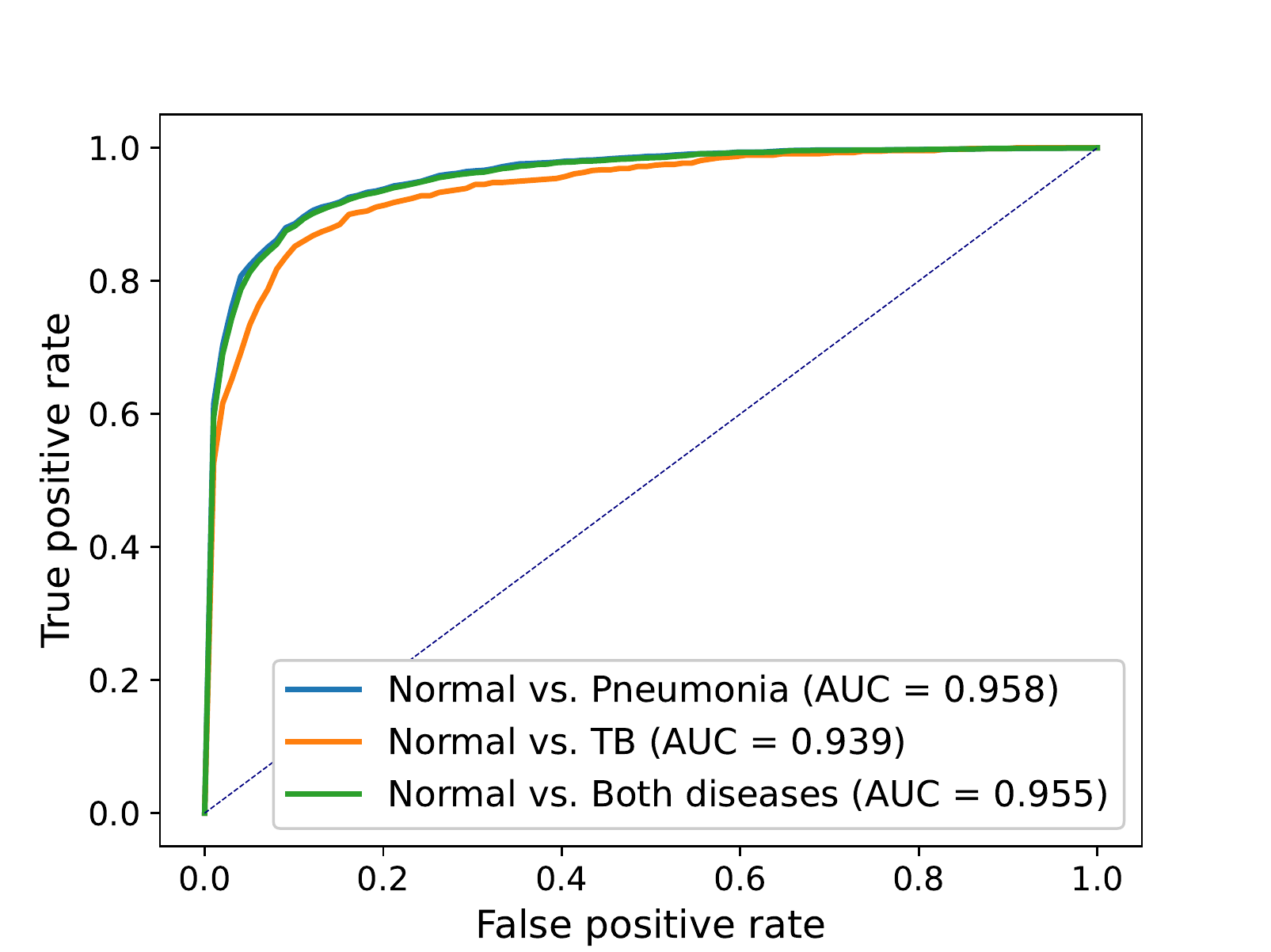}}
	\subfigure[]{\includegraphics[width=0.38\linewidth ]{result_auc_6pro_v2.pdf}}
	\vspace{-0.2cm}
	\caption{\footnotesize ROC curves of VMPR-UAD with different number of projections: (a) two coronal projections, (b) two sagittal projections, (c) four coronal and axial projections, and (d) six coronal, sagittal, and axial projections (proposed number of projections). The mean was obtained from fivefold Monte Carlo validation. }
	\label{fig:projection_num}
\end{figure*}

\subsubsection{Performance of VMPR-UAD according to number of projections}
\label{comp_projnum}

\,\,\,\,\,\,\,\, For the proposed VMPR-UAD, six projections were obtained from the sagittal, coronal, and axial planes separately for the left and right lung regions. To confirm the effectiveness of this configuration, the performance using projections on one plane (i.e., two projected images for the lung regions) and using both coronal and axial views (i.e., four projected images for the lung regions) were compared, obtaining the results listed in Table \mbox{\ref{tab:number_of_projection}} and shown in Fig. \mbox{\ref{fig:projection_num}}. The six projections provide the highest performance regarding AUC, accuracy, sensitivity, specificity, precision, and F1 score, demonstrating the effectiveness of diagnosis considering the three projection planes in VMPR-UAD.

\begin{figure}[hbt!]
	\vskip -15pt
	\centering
	\subfigure[]{\includegraphics[width=0.38\linewidth ]{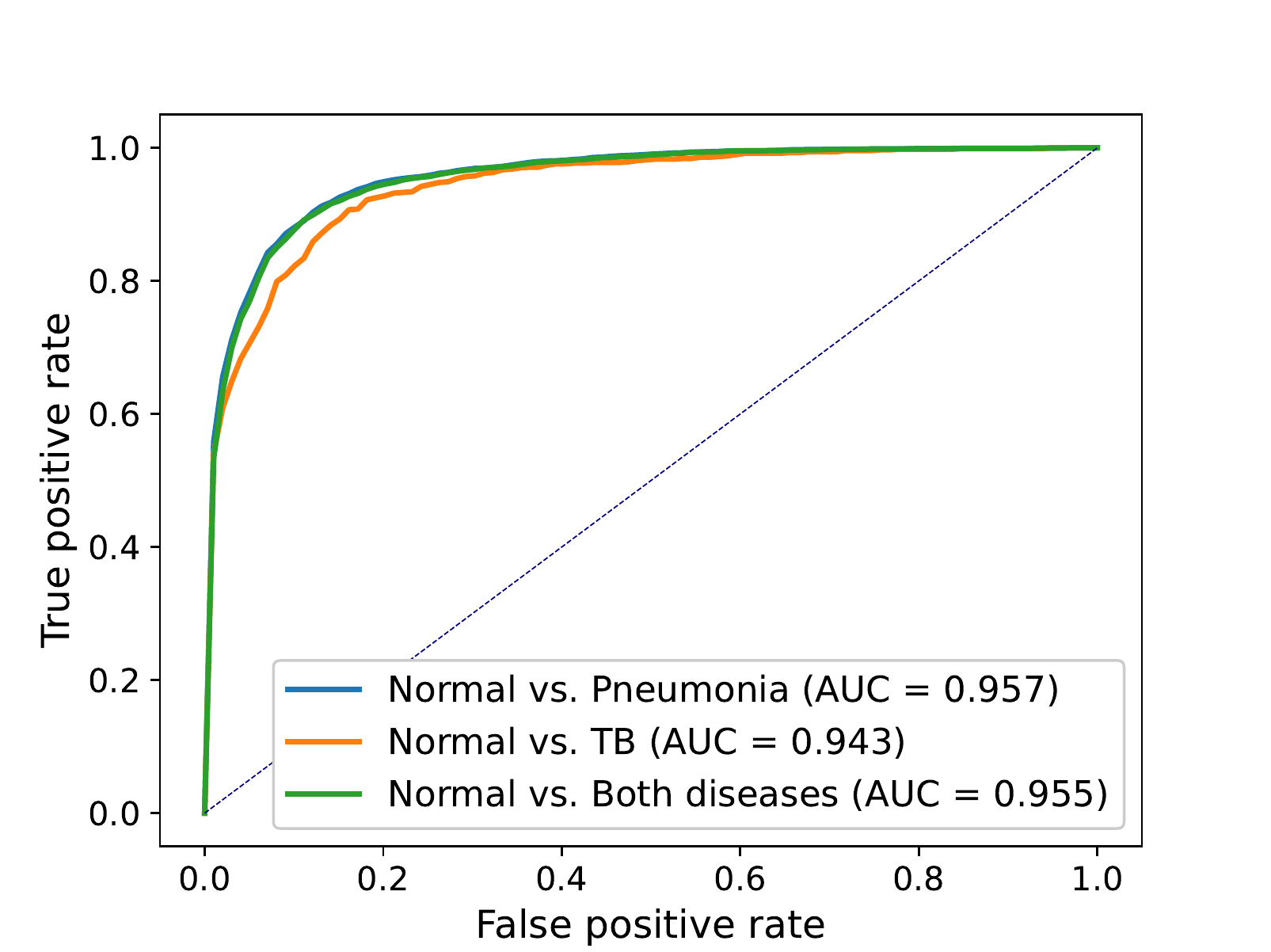}}
	\subfigure[]{\includegraphics[width=0.38\linewidth ]{result_auc_6pro_v2.pdf}}
	\vskip -8pt
	\caption{\footnotesize ROC curves obtained from (a) AIP (baseline) and (b) proposed MIP. The mean was obtained from fivefold Monte Carlo validation. }
	\label{fig:projection_sum_max}
\end{figure}

\subsubsection{Performance according to projection method: Average intensity projection and proposed MIP}
\label{comp_projtype}

\,\,\,\,\,\,\,\, The proposed VMPR-UAD performs MIP (Section \ref{stage1-VMPR-UAD}), whose validity and superiority were confirmed by evaluating the use of MIP and another representative method, average intensity projection (AIP)  \citep{liseo2019dosimetric}. The evaluation results are shown in Fig. \ref{fig:projection_sum_max} and listed in Table \mbox{\ref{tab:sum_vs_max}}. For AIP evaluation, it only replaced MIP while maintaining all the other VMPR-UAD configurations. MIP (Fig. \mbox{\ref{fig:projection_sum_max}}(b)) provides an average AUC of 0.959  (0.965, 0.948, and 0.963 for different detections), whereas AIP (Fig. \mbox{\ref{fig:projection_sum_max}}(a)) provides an average AUC of 0.952 (0.957, 0.943, and 0.955 for different detections), confirming the superiority of the proposed MIP. All the statistical metrics listed in Table \ref{tab:sum_vs_max} show that MIP outperforms AIP, explaining why we adopted MIP in the proposed VMPR-UAD.

\begin{table}[hbt!]
	\caption{\footnotesize {Statistical results obtained by using MIP and AIP. The mean and standard deviation were obtained from fivefold Monte Carlo validation.}}
	\footnotesize
	\centering
	{
		\resizebox{.8\linewidth}{!}{
	    	\begin{tabular}{ccccccc}
				\toprule
			    Projection & Label & Accuracy & Sensitivity &Specificity& Precision& F1 score \\
				
				\midrule
				
			    \multirow{3}{*}{AIP}
				&Pneumonia  & 
				0.895 $\pm$ 0.01 & 
				0.893 $\pm$ 0.01 & 
				0.897 $\pm$ 0.01 & 
				0.896 $\pm$ 0.01 & 
				0.895 $\pm$ 0.01 \\
				
				&Tuberculosis  & 
				0.879 $\pm$ 0.01 &
				0.885 $\pm$ 0.02 &
				0.873 $\pm$ 0.03 &
				0.875 $\pm$ 0.02 &
				0.880 $\pm$ 0.01 \\
				
				&Both diseases &
			    0.892 $\pm$ 0.01 & 
				0.893 $\pm$ 0.01 & 
        		0.892 $\pm$ 0.01 &
        		0.892 $\pm$ 0.01 & 
        		0.893 $\pm$ 0.01 \\

				\midrule
			    \multirow{3}{*}{MIP}
				&Pneumonia  & 
				\textbf{0.907 $\pm$ 0.01} &
				\textbf{0.902 $\pm$ 0.01} &
				\textbf{0.913 $\pm$ 0.01} & 
				\textbf{0.912 $\pm$ 0.01} & 
				\textbf{0.907 $\pm$ 0.01}  \\
				
				&Tuberculosis  & 
				\textbf{0.893 $\pm$ 0.02} &
				\textbf{0.889 $\pm$ 0.01} & 
				\textbf{0.897 $\pm$ 0.03} & 
				\textbf{0.897 $\pm$ 0.03} & 
				\textbf{0.893 $\pm$ 0.02}  \\
				
				&Both diseases &
			    \textbf{0.904 $\pm$ 0.01}& 
				\textbf{0.899 $\pm$ 0.01}&
				\textbf{0.910 $\pm$ 0.01} &
				\textbf{0.909 $\pm$ 0.01} &
				\textbf{0.904 $\pm$ 0.01} \\
				
				\bottomrule
			\end{tabular}

		}
	}
	\label{tab:sum_vs_max}
\end{table}

\begin{figure*}[hbt!]
	\vskip -5pt
	\centering
	
	\subfigure[]{\includegraphics[width=.32\linewidth, height=4.2cm ]{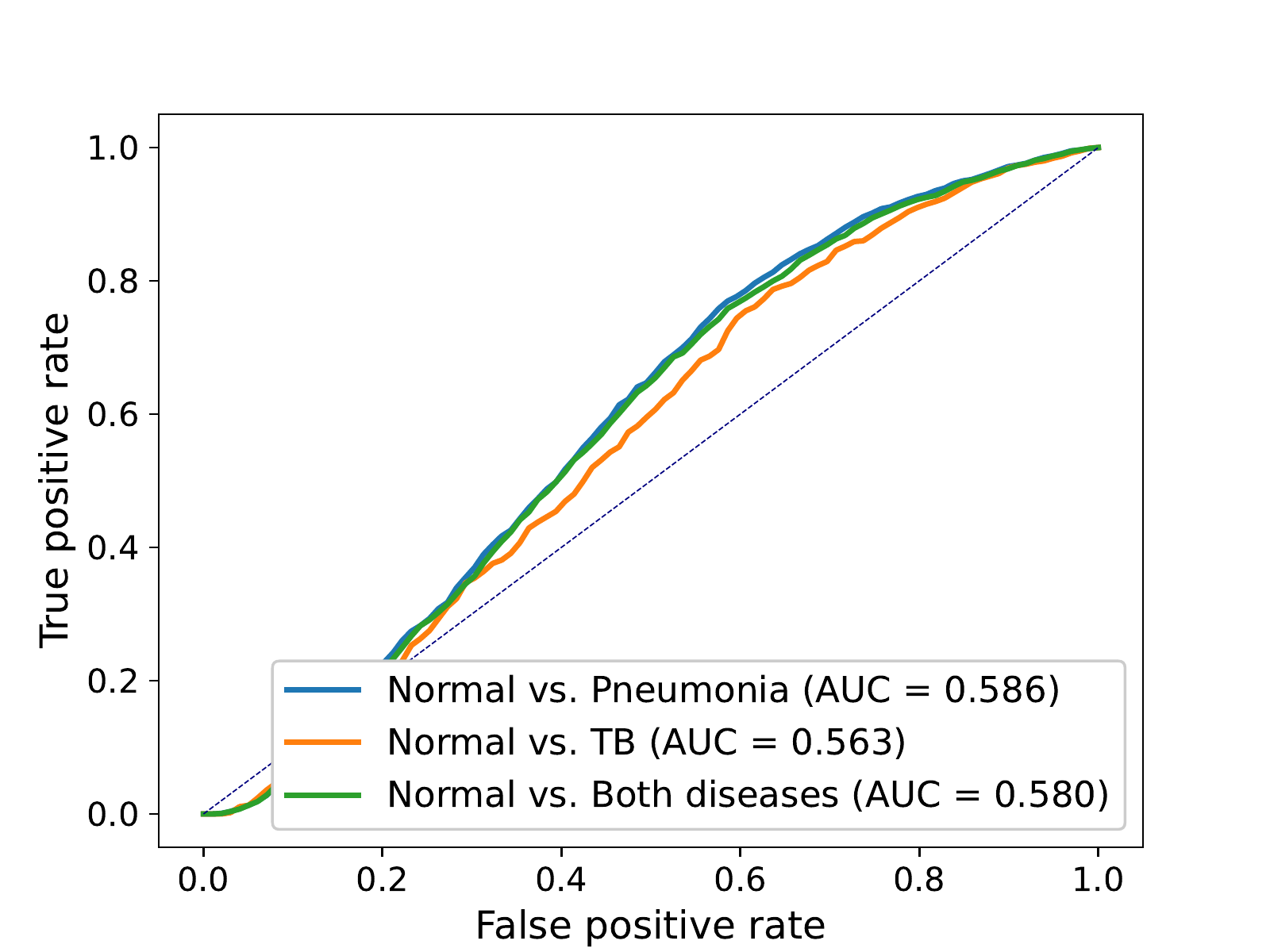}}
	\subfigure[]{\includegraphics[width=.32\linewidth, height=4.2cm ]{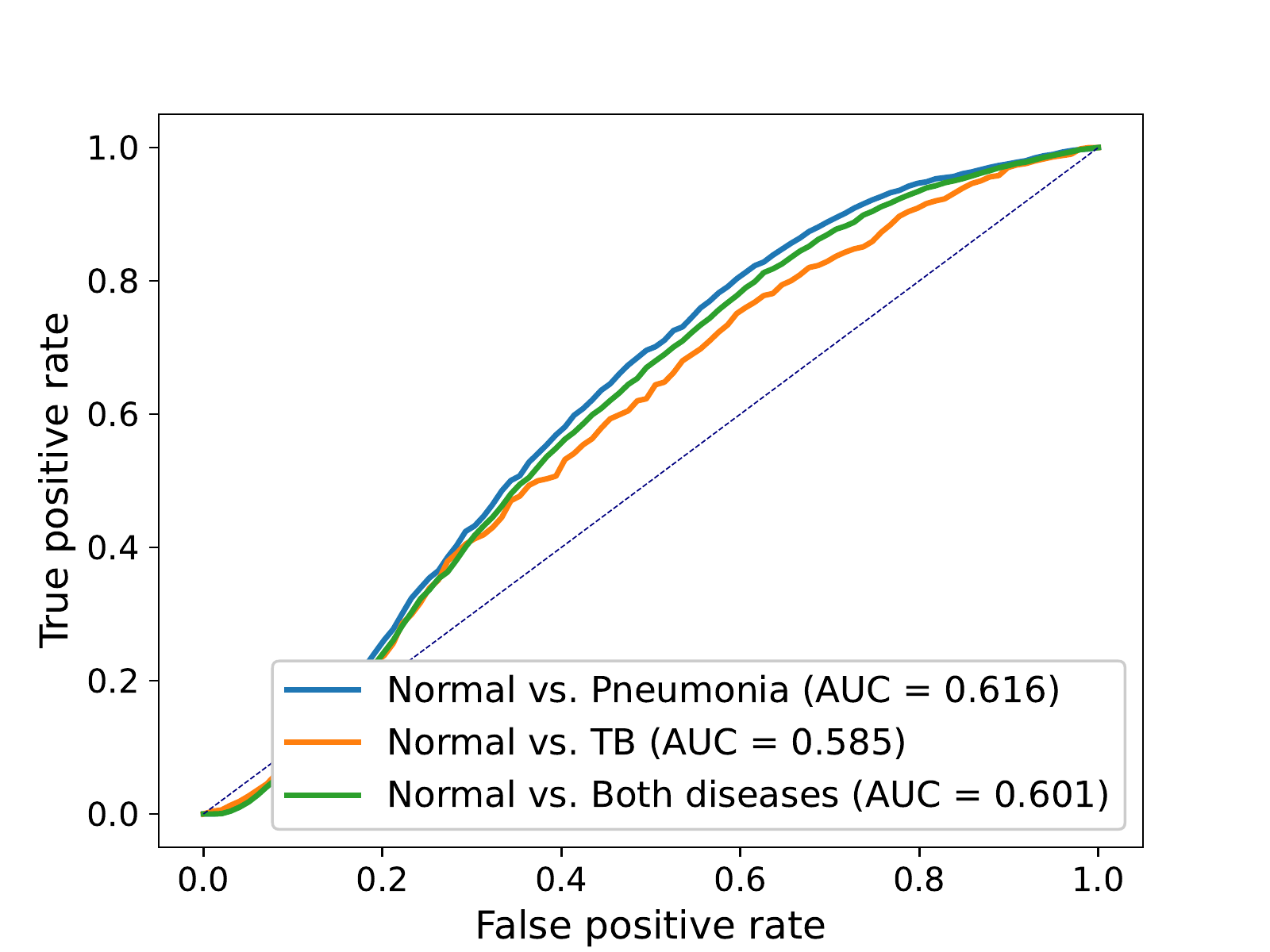}}
	\subfigure[]{\includegraphics[width=.32\linewidth, height=4.2cm ]{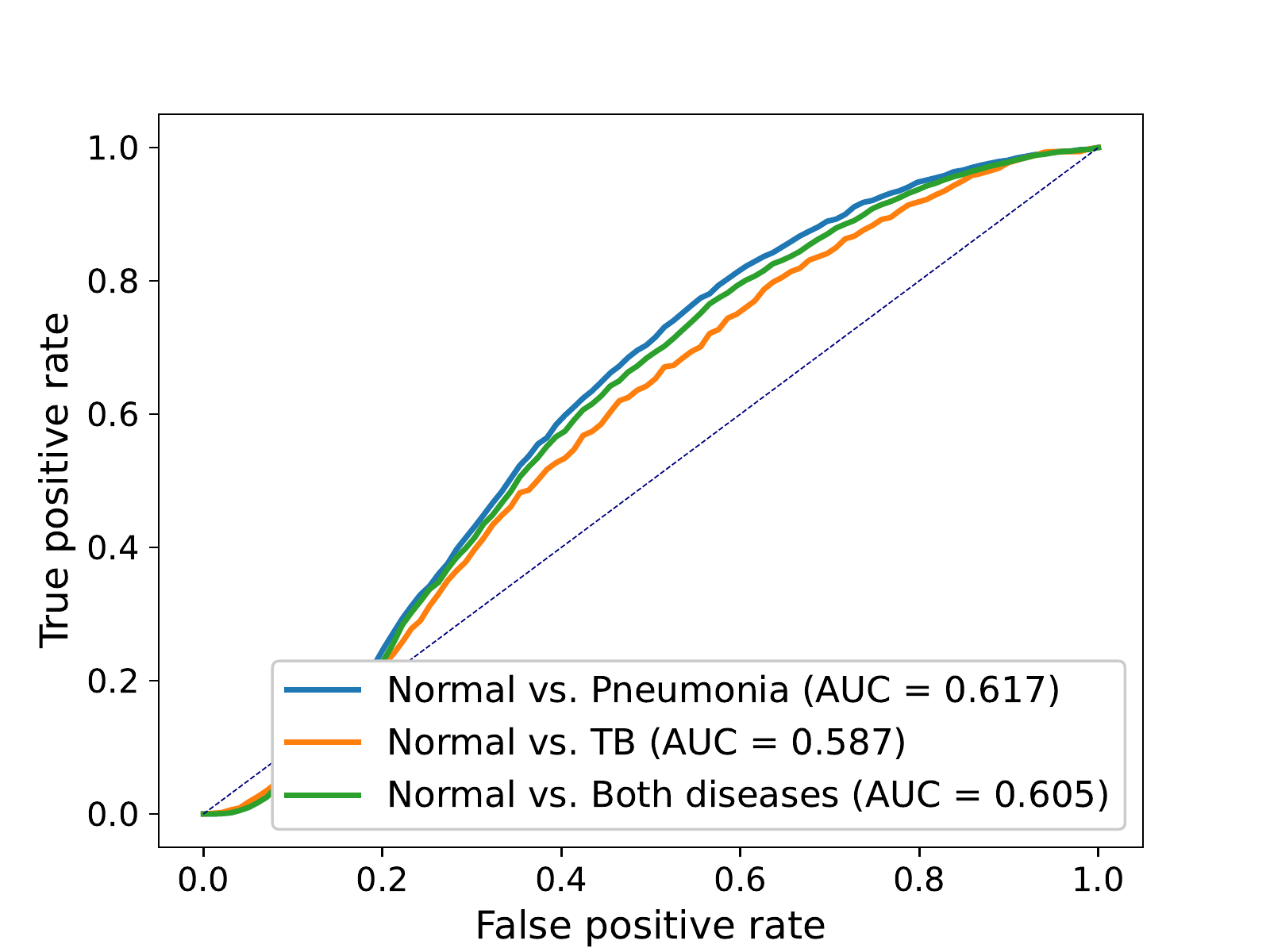}}
	
	\subfigure[]{\includegraphics[width=.32\linewidth, height=4.2cm ]{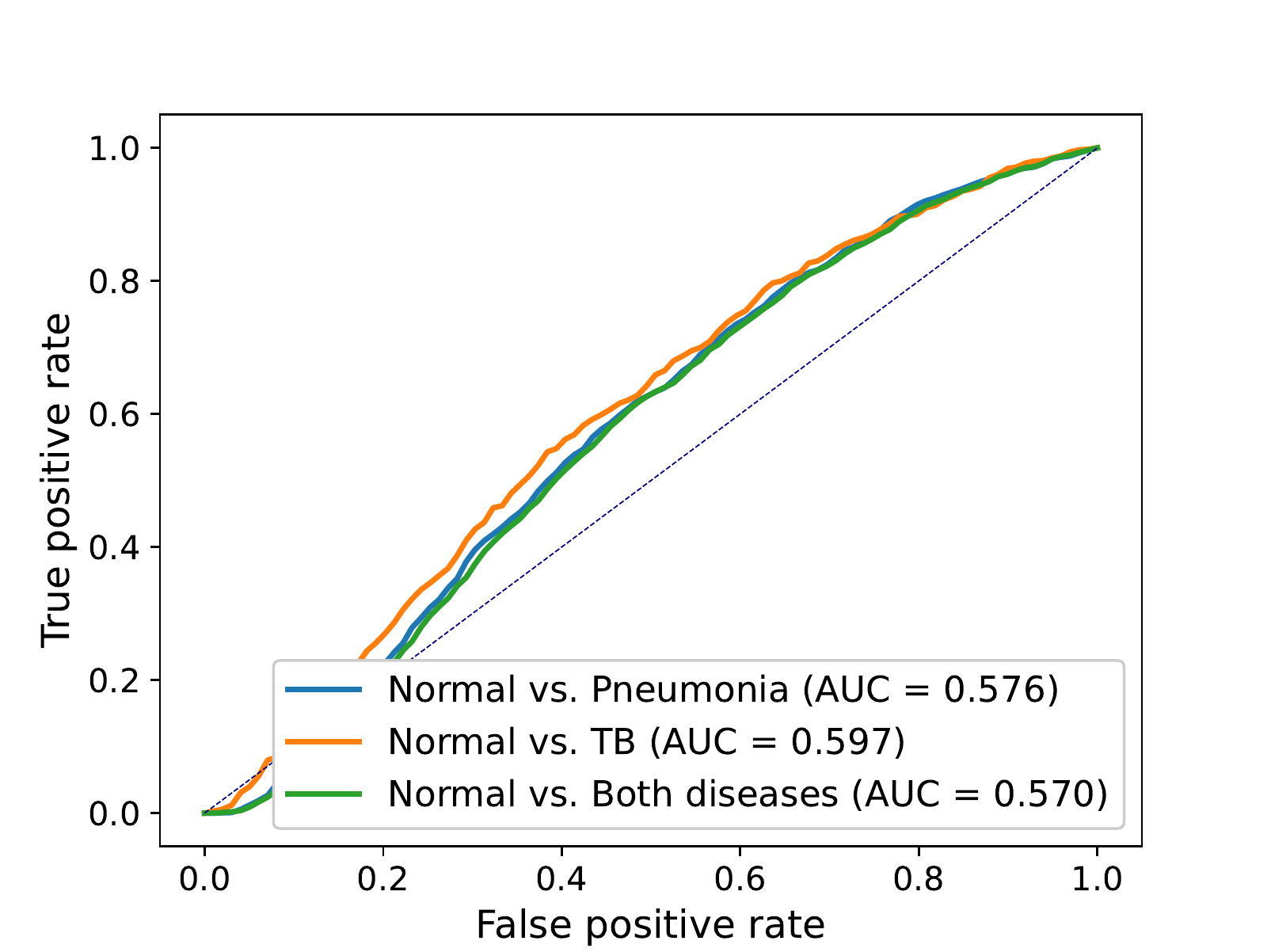}}
	\subfigure[]{\includegraphics[width=.32\linewidth, height=4.2cm ]{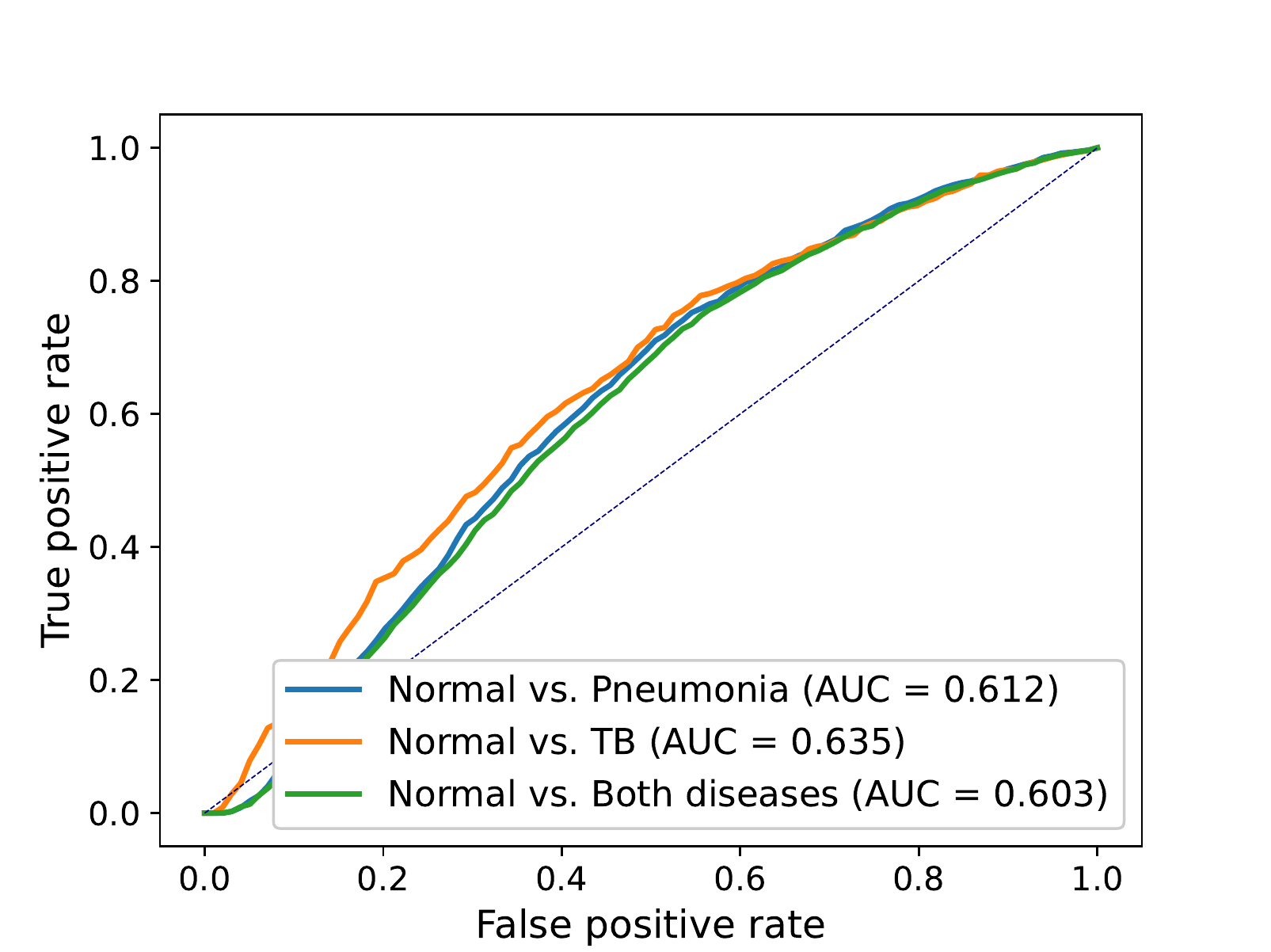}}
	\subfigure[]{\includegraphics[width=.32\linewidth, height=4.2cm ]{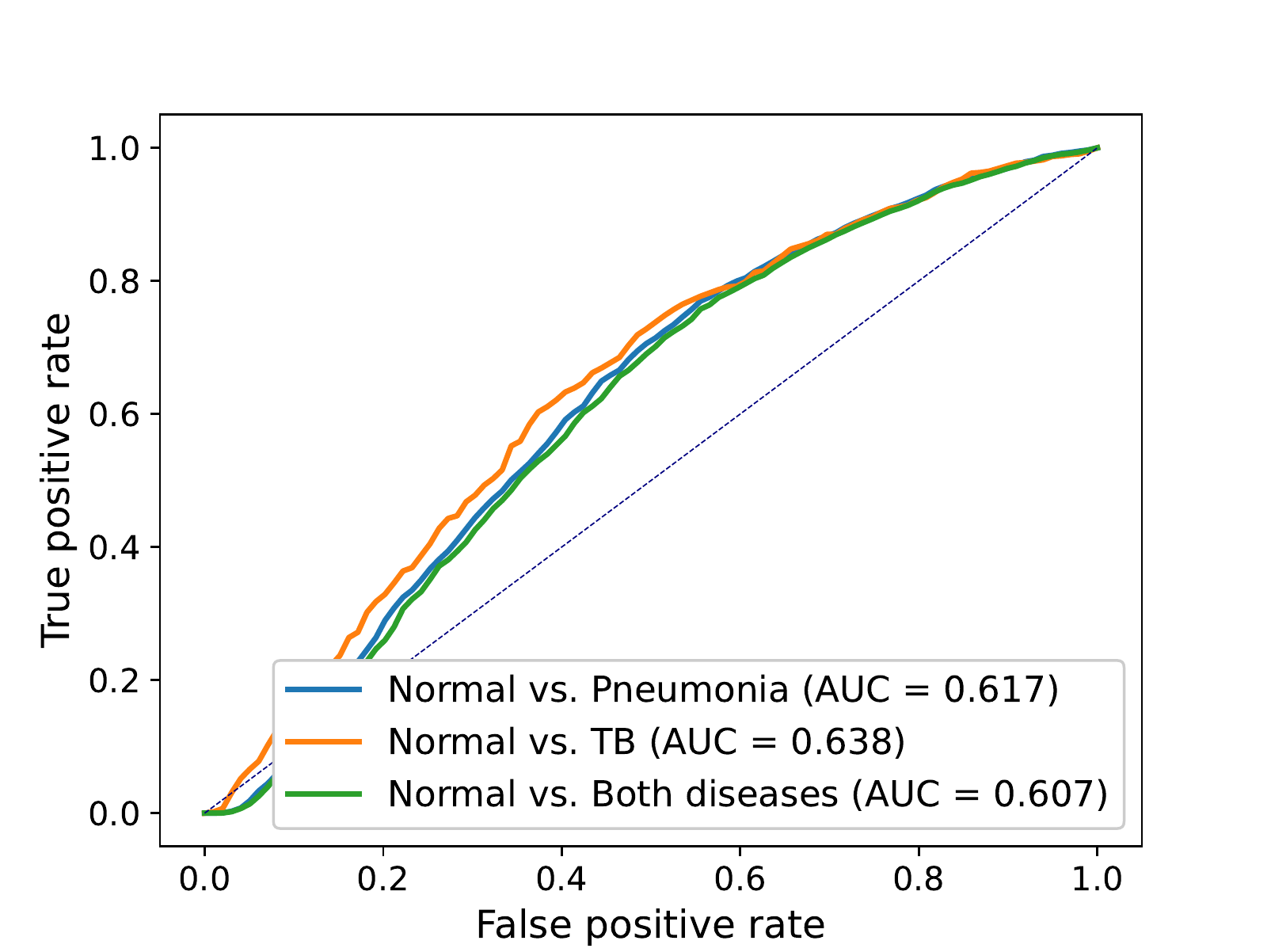}}

	\vspace{-0.2cm}
	\caption{\footnotesize ROC curves of VMPR-UAD using different number and types of projections without segmentation: AIP for (a) coronal projection (two projections), (b) coronal and axial projections (four projections), and (c) coronal, sagittal, and axial projections (six projections) and MIP for (d) coronal projection (two projections), (e) coronal and axial projections (four projections), and (f) coronal, sagittal, and axial projections (six projections). The mean was obtained from fivefold Monte Carlo validation. }
	\label{fig:non_seg_sum_max}
\end{figure*}
 
\subsubsection{Performance with and without segmentation}
\label{comp_seg}

\,\,\,\,\,\,\,\, The proposed VMPR-UAD performed each projection after segmentation of the 3D lung regions. To verify the effectiveness of segmentation, we performed anomaly detection using projection with and without segmentation while maintaining all the other VMPR-UAD configurations, obtaining the results shown in Fig. \mbox{\ref{fig:non_seg_sum_max}}.

After applying different numbers of projections and different types of projections (see Sections \ref{comp_projnum} and \ref{comp_projtype}), we confirmed that the proposed method considering segmentation achieves the highest performance for all the cases. Without segmentation, the anomaly detection provides an approximate AUC of 0.6, as shown in Fig. \mbox{\ref{fig:non_seg_sum_max}}, whereas the complete VMPR-UAD considering segmentation achieves an approximate AUC of 0.9, as shown in Fig. \mbox{\ref{fig:sl_comp}}. These results indicate a performance improvement of approximately 50\% by considering segmentation, verifying its importance in the proposed VMPR-UAD.

\begin{figure*}[hbt!]
	\vskip -5pt
	\centering
	\subfigure[]{\includegraphics[width=0.32\linewidth, 
	height=4.2cm ]{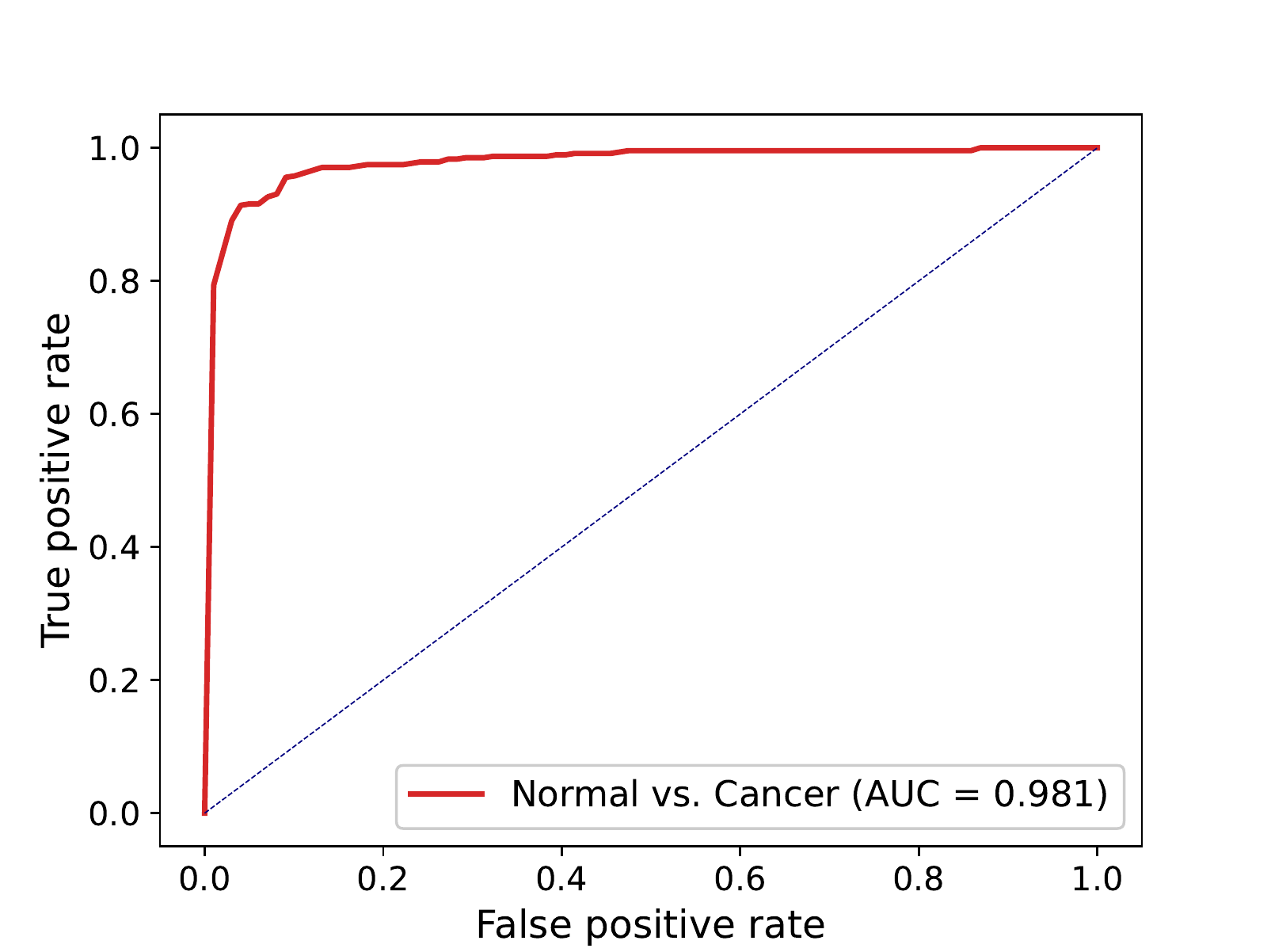}}
	\subfigure[]{\includegraphics[width=0.32\linewidth, 
	height=4.2cm ]{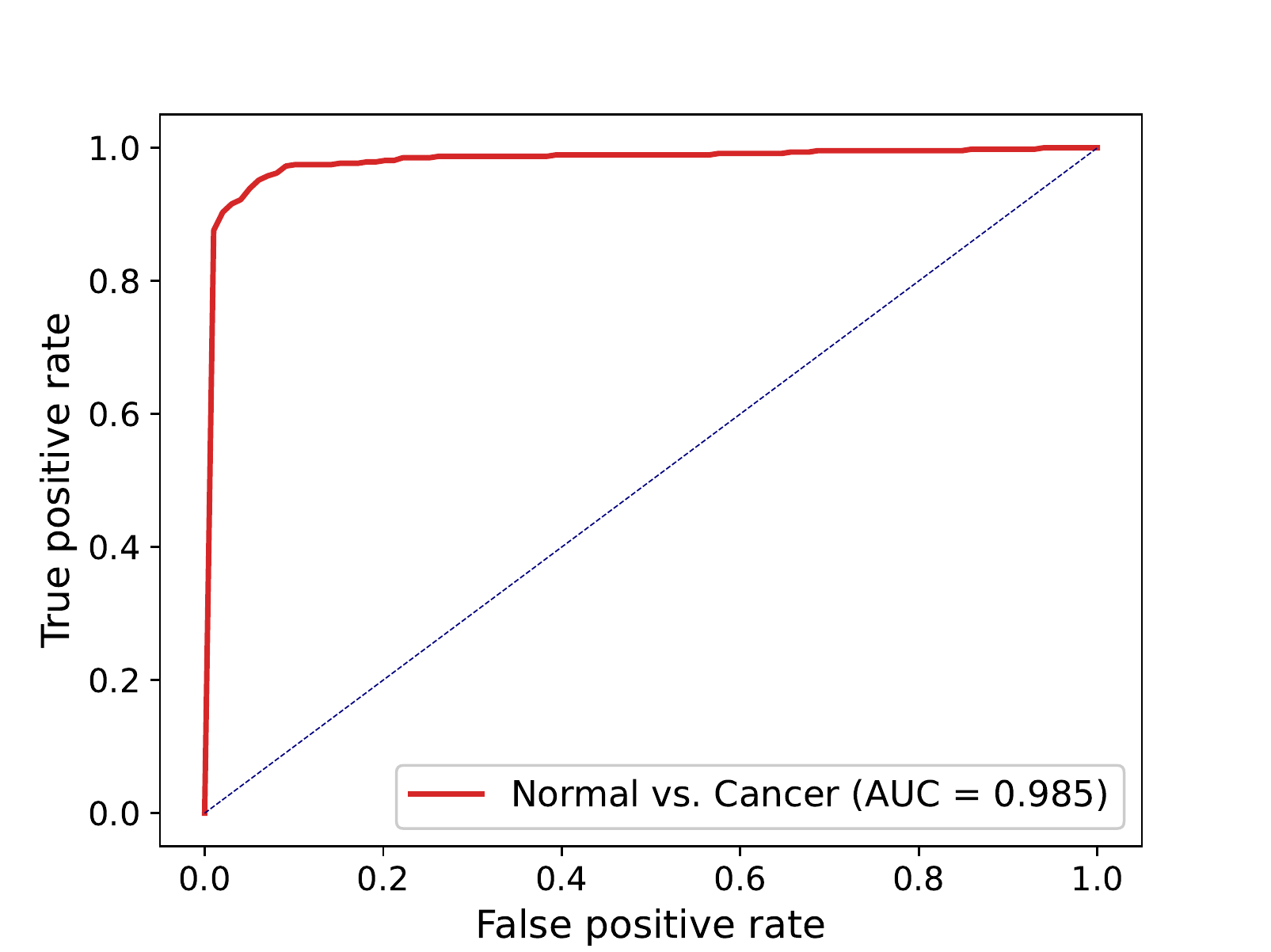}}
	\subfigure[]{\includegraphics[width=0.32\linewidth, 
	height=4.2cm ]{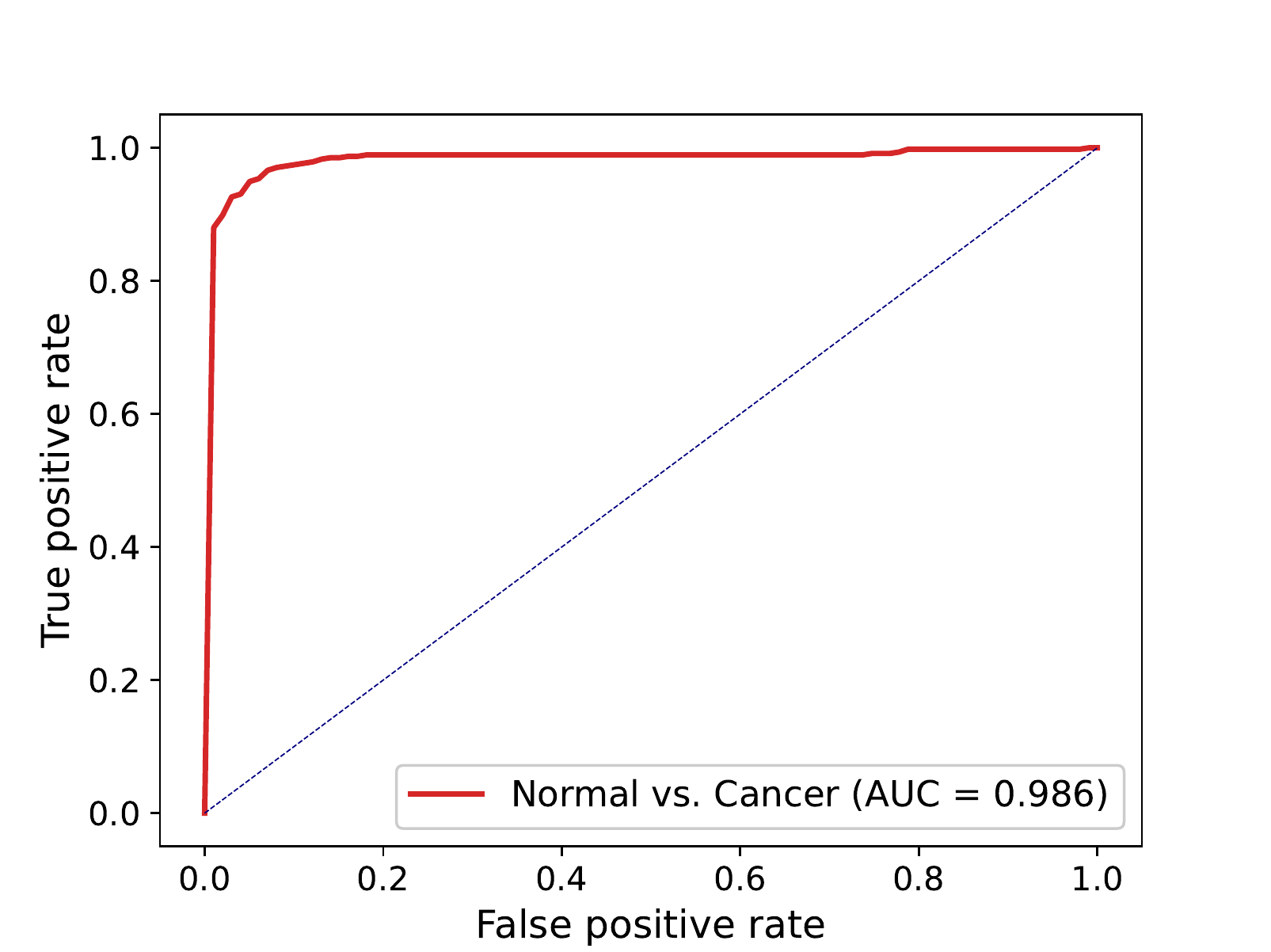}}
	\vspace{-0.2cm}
	\caption{\footnotesize ROC curves and AUC for external validation with different numbers of projections: (a) coronal projection (two projections), (b) coronal and axial projections (four projections), and (c) coronal, sagittal, and axial projections (six projections). The mean was obtained from fivefold Monte Carlo validation.  }
	\label{fig:external_roc}
\end{figure*}

\begin{figure*}[hbt!]
	\vskip -5pt
	\centering
	\subfigure[]{\includegraphics[width=0.95\linewidth]{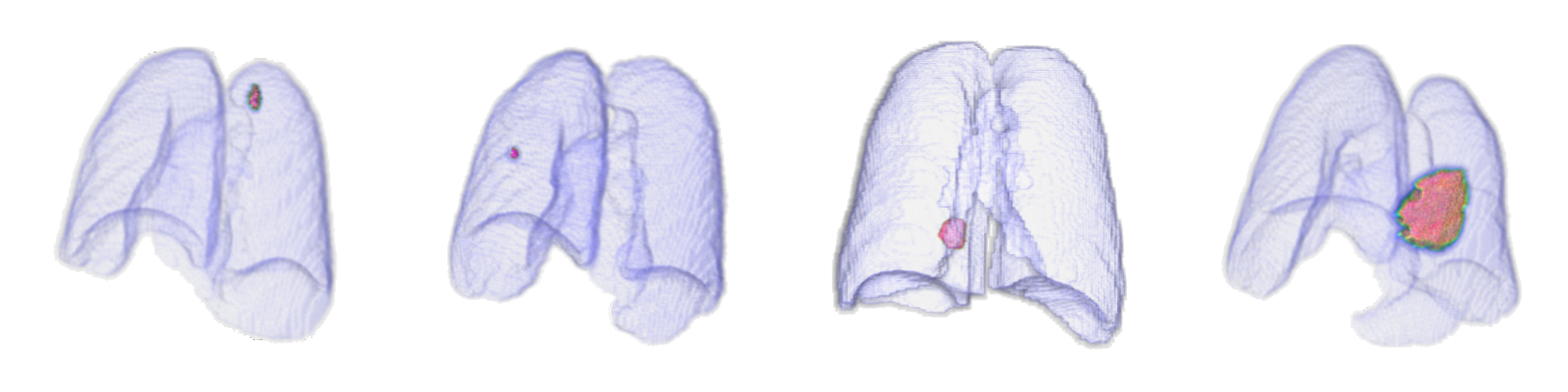}}
	\vskip -4pt
	\subfigure[]{\includegraphics[width=0.95\linewidth]{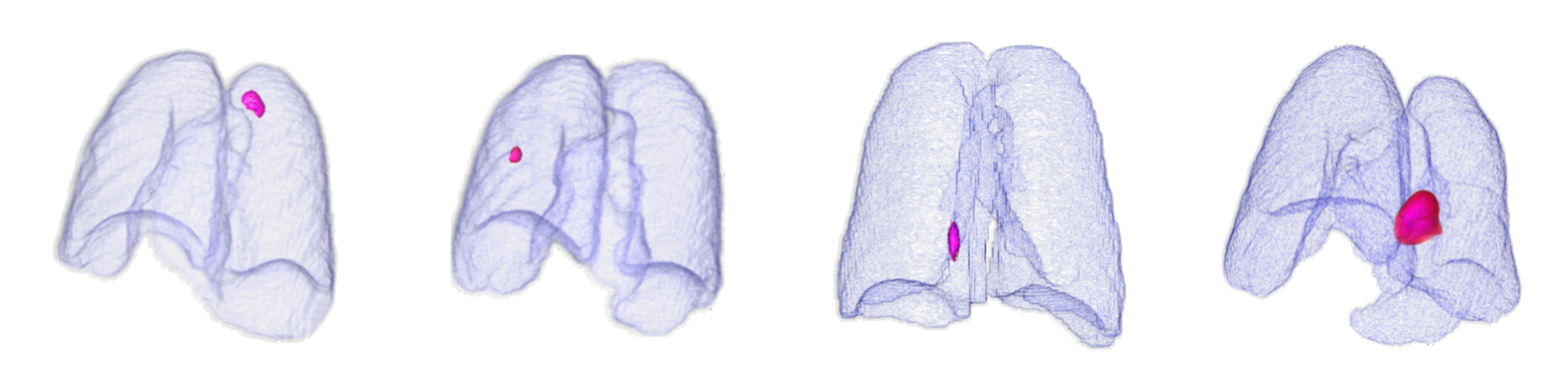}}
	\vspace{-0.2cm}
	\caption{\footnotesize Examples of 3D anomaly maps for lung cancer detection. (a) Ground truth and (b) 3D anomaly map generated by VMPR-UAD.}
	\label{fig:result_gt_cam}
\end{figure*}

\subsection{External validation using public lung cancer CT dataset}
\,\,\,\,\,\,\,\, The experimental results presented thus far (Sections \ref{comp_sl}, \ref{comp_projnum}, \ref{comp_projtype}, and \ref{comp_seg}) show verification using SMC internal data (i.e., anomaly detection for pneumonia, tuberculosis, and both diseases). In this section, we further confirm the superiority of the proposed VMPR-UAD by demonstrating its high performance on external data, that is, the MSD dataset with 95 3D stacks showing lung cancer   \citep{antonelli2021medical}). The anomaly detection results of VMPR-UAD are shown as ROC curves with AUC in Fig. \ref{fig:external_roc}. VMPR-UAD provides high anomaly detection performance (at least 0.98) for different numbers of projections, with a higher diagnostic performance achieved when using more projections. As this trend is the same as that obtained from internal data, we can infer that VMPR-UAD consistently provides high diagnostic performance regardless of the CT dataset.

\begin{figure*}[t]
\centerline{\includegraphics[width=0.95\textwidth, height=11cm]{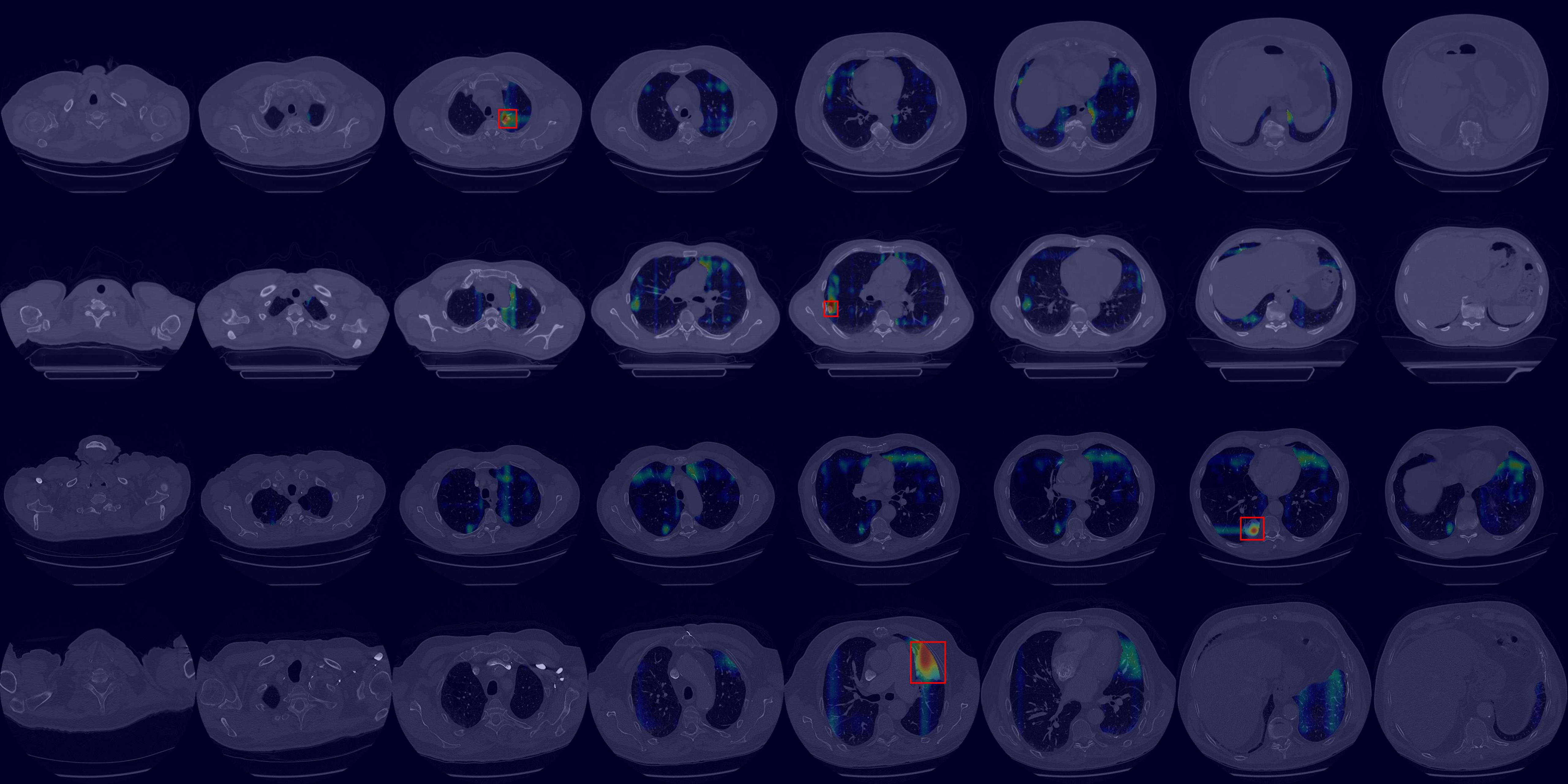}} 
\caption{Slice-level examples of anomaly maps for lung cancer localization. The anomaly maps generated by VMPR-UAD correctly indicate the cancer location in the red boxes.}
\label{fig:result_cam_cancer}
\end{figure*}

Finally, we evaluated the 3D abnormal (lesion) localization performance of the proposed VMPR-UAD. Of the 63 cancer cases in the MSD dataset with ground-truth annotations available, we excluded two cases (cases 38 and 96) of incorrect annotation or showing other diseases. We calculated whether the cancer area predicted by VMPR-UAD (binarized at a certain high-probability threshold) overlapped with the annotated cancer area. As a result, 57 of the 61 cases showed overlapping, demonstrating that VMPR-UAD can localize 3D lung anomalies (cancer in this case) with an accuracy of 93$\%$. Some localization examples in 3D data are shown in Fig. \mbox{\ref{fig:result_gt_cam}}. The red points in Fig. \mbox{\ref{fig:result_gt_cam}}(b) show the 3D cancer locations that the proposed method estimates with the highest confidence (i.e., location of highest pixel value in the 3D anomaly map). The red points in Fig. \mbox{\ref{fig:result_gt_cam}}(a) show the ground-truth cancer location. The ground truth and prediction shown in Fig. \mbox{\ref{fig:result_gt_cam}} confirm that VMPR-UAD correctly finds the lung anomaly 3D region. More detailed visualization results are available in the Supplementary Material. The prediction consistency can also be observed in 2D slices, as shown in Fig. \ref{fig:result_cam_cancer}, where our anomaly localization map indicates correct cancer regions. Hence, the proposed VMPR-UAD can automatically localize or segment lesions without requiring any lesion information (i.e., using only CT slices from healthy subjects) for training.

\section{Conclusion}
\,\,\,\,\,\,\,\, We propose VMPR-UAD using only LDCT data from healthy subjects for training to automatically indicate the presence or absence of anomalies and their location corresponding to lesions shown in 3D stacks of patients with lung diseases. The proposed method provides a higher performance than existing SL-based methods, overcoming difficulties in implementing existing SL-based CT computer-aided diagnosis in practice owing to the annotation burden and the clinical inefficiency given the low performance of existing UAD methods. In particular, SL-based methods cannot guarantee anomaly detection in unseen/untrained data (e.g., a new infectious disease with insufficient training data such as a COVID-19 variant). Thus, we expect VMPR-UAD to facilitate diagnosis of an emerging pandemic by supporting first-line testing of patients infected by an unknown infectious lung disease while avoiding annotation burdens. In addition, the appropriate 3D lesion localization suggests the applicability of VMPR-UAD as an unsupervised-learning network pretraining tool, which can be fine-tuned for accurate disease localization and segmentation by using only a small amount of training data from disease cases. Thus, diverse complementary methods can be derived from our development.  
 
\section*{CRediT author statement}
\,\,\,\,\,\,\,\, Kyung-Su Kim led the design, writing, and analysis, Seong Je Oh performed the key analysis and assisted with writing, Ju Hwan Lee assisted with the experimental analysis, and Myung Jin Chung provided LDCT data and validated our results.
 
\section*{Acknowledgements}
\,\,\,\,\,\,\,\, This work was supported by the National Research Foundation of Korea (NRF) grant funded by the Korean government (MSIT) (2021R1F1A106153511). This work was also supported by the Korea Medical Device Development Fund grant funded by the Korean government (Ministry of Science and ICT, Ministry of Trade, Industry and Energy, Ministry of Health $\&$ Welfare, Ministry of Food and Drug Safety) (202011B08-02, KMDF$\_$PR$\_$20200901$\_$0014-2021-02). This work was also supported by the Technology Innovation Program (20014111) funded by the Ministry of Trade, Industry $\&$ Energy (MOTIE, Korea). This work was also supported by the Future Medicine 20*30 Project of the Samsung Medical Center (SMX1210791).

\clearpage 
\bibliographystyle{unsrtnat}
\bibliography{refs.bib}
\clearpage 

\appendix
\section{Supplemental materials}

\begin{figure*}[hbt!]
	\vskip -5pt
	\centering
	\subfigure[]{\includegraphics[width=0.95\linewidth]{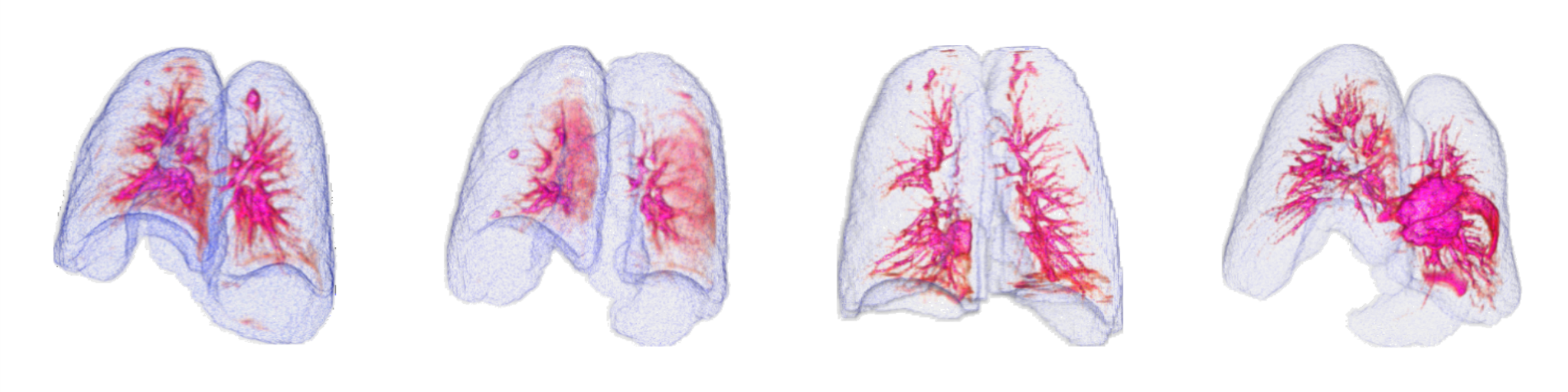}}
	\vskip -4pt
	
	\subfigure[]{\includegraphics[width=0.95\linewidth]{gt_results.pdf}}
	\vskip -4pt

	\subfigure[]{\includegraphics[width=0.95\linewidth]{cam_results.pdf}}
	\vspace{-0.2cm}
	\caption{\footnotesize Examples of 3D anomaly maps for lung cancer case. (a) Segmented lung, (b) Ground-truth cancer location, and (c) 3D anomaly map generated by VMPR-UAD. The proposed VMPR-UAD correctly localizes the anomalies related to lung cancer.}
	\label{fig:3d_cancer}
\end{figure*}

\begin{figure*}[t]
	\vskip -5pt
	\centering
	\subfigure[]{\includegraphics[width=0.9\linewidth]{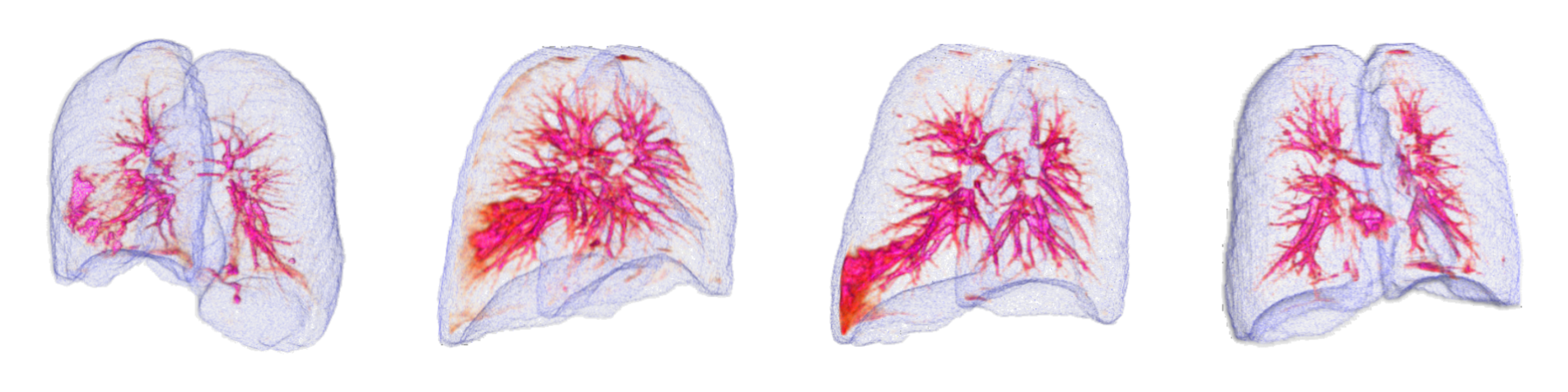}}
	\vskip -4pt
	
	\subfigure[]{\includegraphics[width=0.95\linewidth]{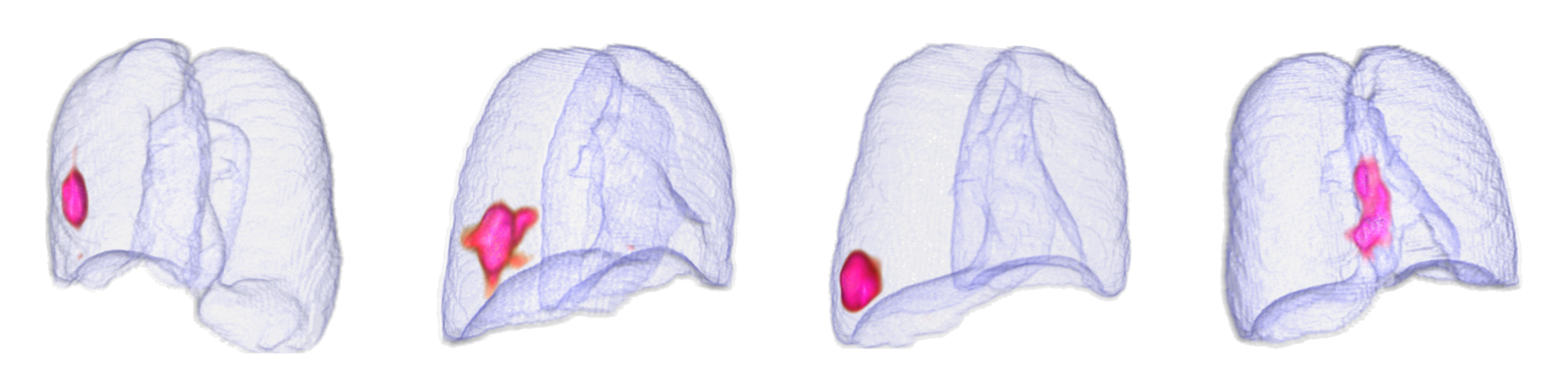}}
	\vspace{-0.2cm}
	
    \subfigure[]{\includegraphics[width=0.95\linewidth]{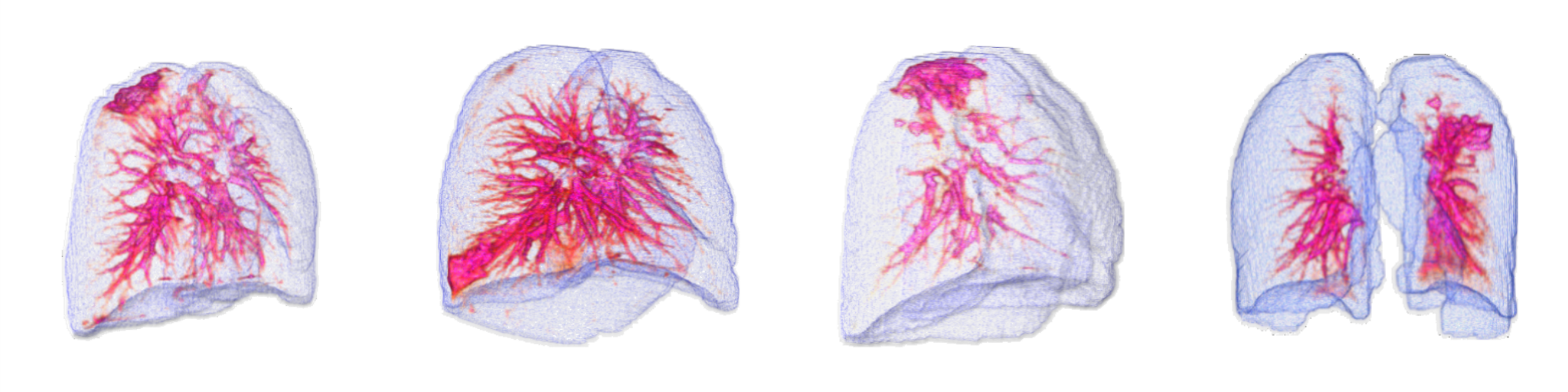}}
	\vskip -4pt
	
	\subfigure[]{\includegraphics[width=0.95\linewidth]{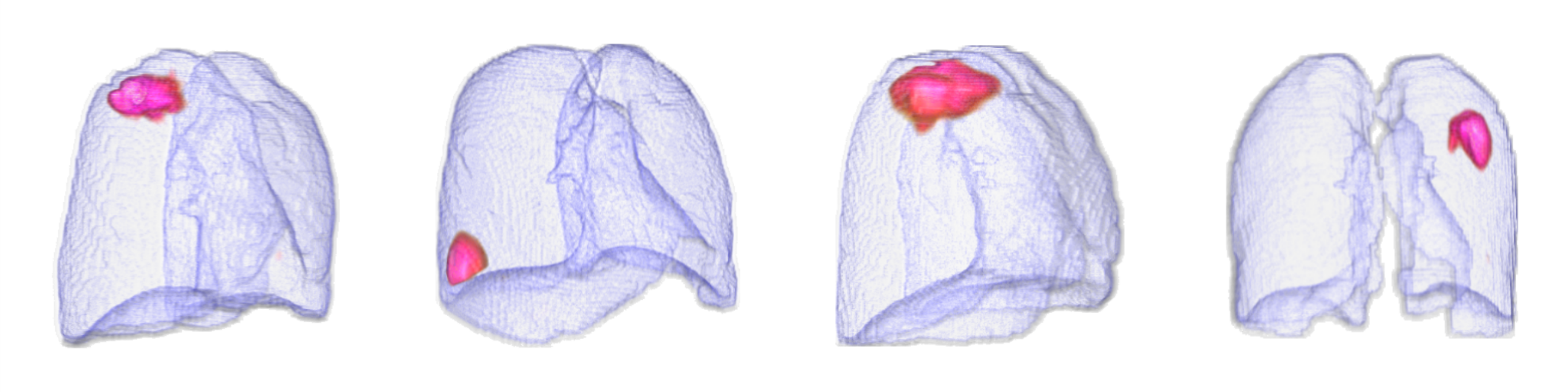}}
	\vspace{-0.2cm}

	\caption{\footnotesize Examples of 3D anomaly maps for lung pneumonia and tuberculosis cases. (a) Segmented lung and (b) 3D anomaly maps generated by VMPR-UAD for patients with pneumonia. (c) Segmented lung and (d) 3D anomaly maps generated by VMPR-UAD for patients with tuberculosis. The proposed VMPR-UAD correctly localizes anomalies related to pneumonia and tuberculosis.}
	\label{fig:3d_pneumonia_tb}
\end{figure*}

\begin{figure*}[t]
\centerline{\includegraphics[width=0.95\textwidth]{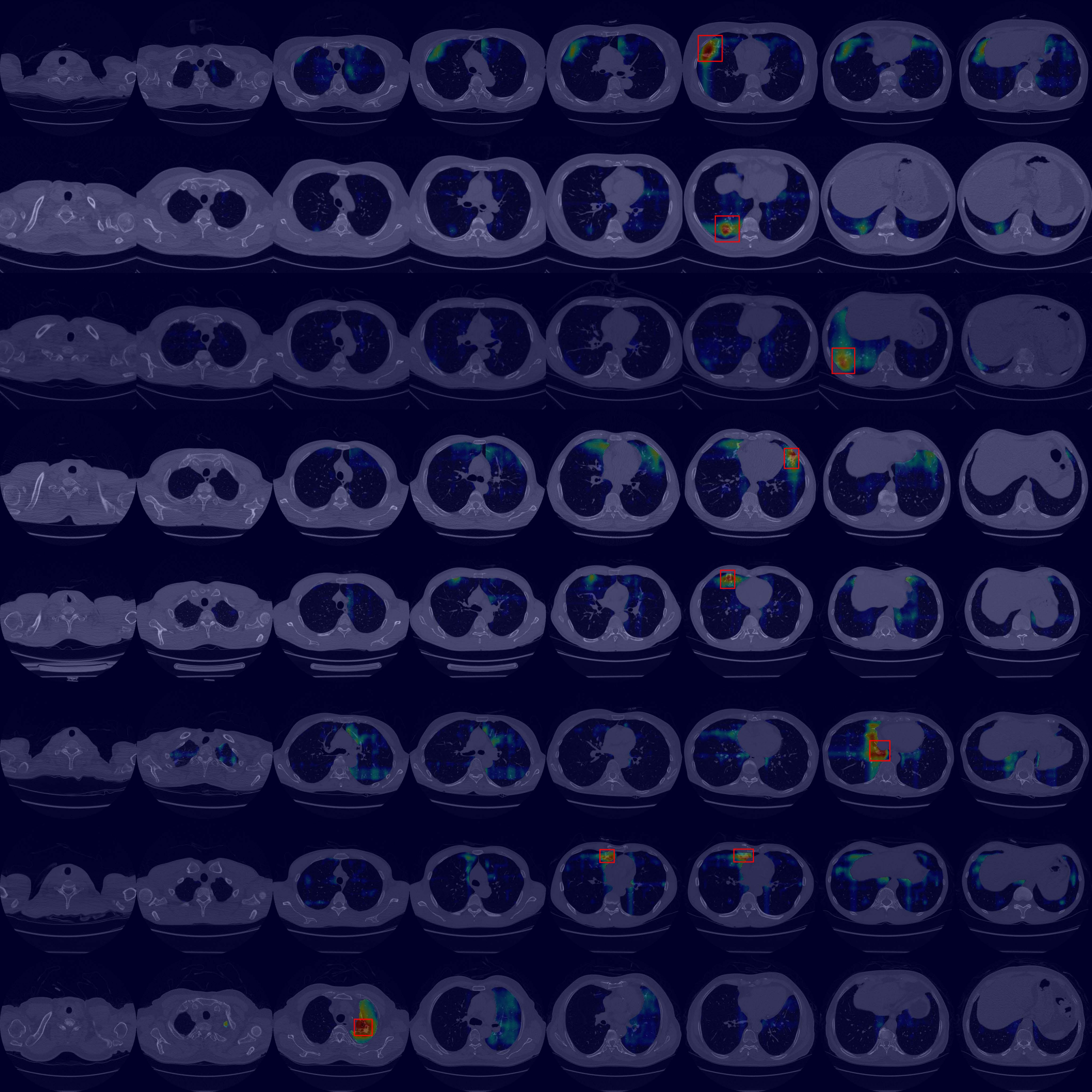}} 
\caption{Slice-level examples of anomaly maps for pneumonia case. The anomaly maps generated by VMPR-UAD correctly indicate the anomaly in the red boxes.}
\label{fig:2d_pneumonia}
\end{figure*}

\begin{figure*}[t]
\centerline{\includegraphics[width=0.95\textwidth]{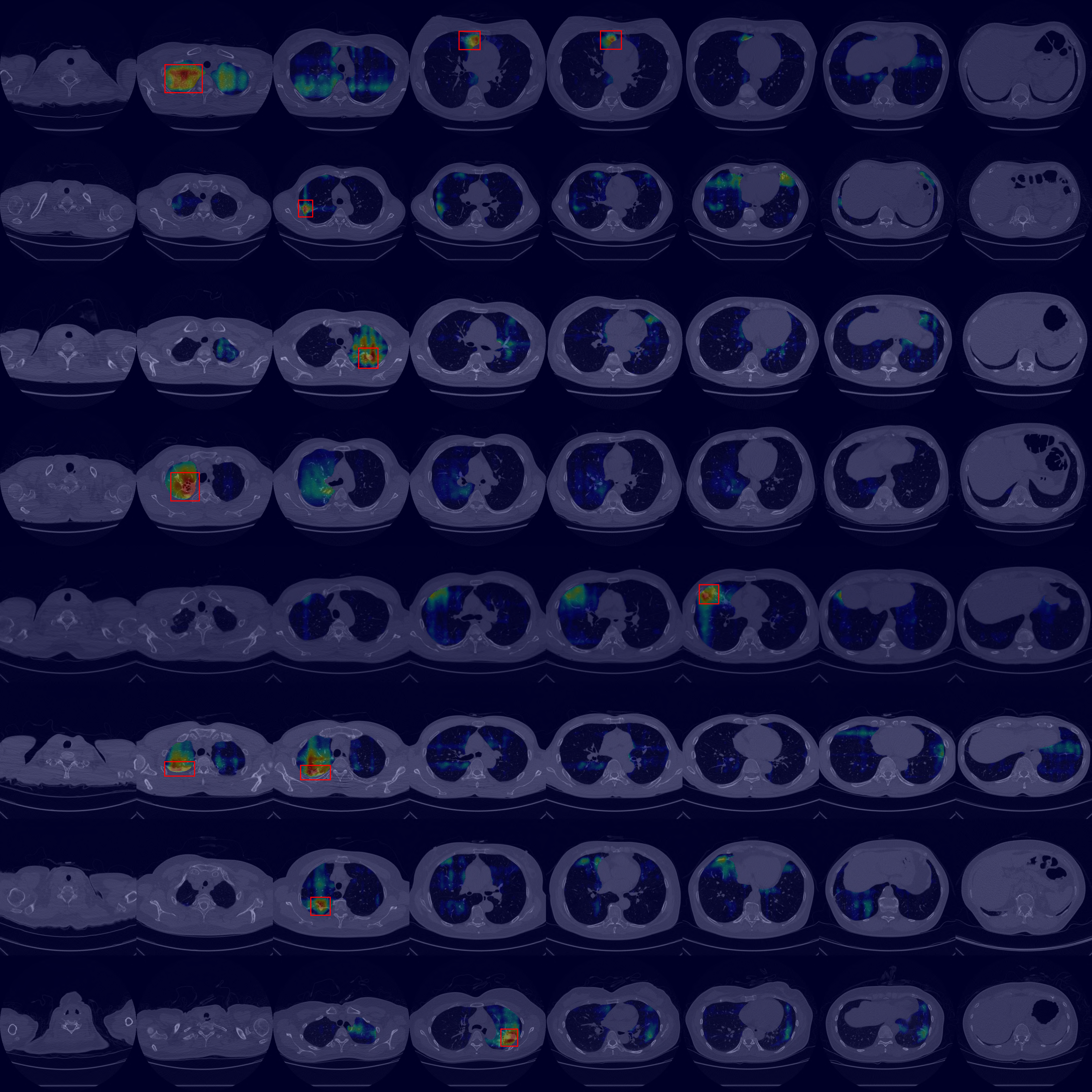}} 
\caption{Slice-level examples of anomaly maps for tuberculosis case. The anomaly maps generated by VMPR-UAD correctly indicate the anomaly in the red boxes.}
\label{fig:2d_tb}
\end{figure*}

\end{document}